\documentclass[aps,prl,showpacs,twocolumn,floats,epsfig,superscriptaddress,pdflatex]{revtex4}
\usepackage{epsfig}
\usepackage{amsmath}
\usepackage{amsfonts}
\usepackage{graphicx}
\usepackage{amssymb}
\usepackage{amsbsy}
\usepackage{xcolor}
\usepackage{ulem}
\usepackage{array}
\usepackage{latexsym}
\usepackage{amsbsy}
\usepackage{epstopdf}
\usepackage{bm}
\usepackage{float}

\newcommand{\bra}[1]{\langle{#1}|}
\newcommand{\ket}[1]{|{#1}\rangle}

\newcommand{\red}{\color{red}}
\newcommand{\blue}{\color{blue}}

\newcommand{\WQC}{Wilczek Quantum Center, School of Physics and
Astronomy, Shanghai Jiao Tong University, Shanghai 200240, China}
\newcommand{\Pittsburgh}{Department of Physics and Astronomy, University of Pittsburgh, Pittsburgh, PA 15260, USA}
\newcommand{\TDLee}{T. D. Lee Institute, Shanghai Jiao Tong University, Shanghai 200240, China}
\newcommand{\SUSTECH}{Shenzhen Institute for Quantum Science and Engineering, Southern University of Science and Technology, 
Shenzhen 518055, China}
\newcommand{\Key}{Key Laboratory of Artificial Structures and Quantum Control, School of Physics and
Astronomy, Shanghai Jiao Tong University, Shanghai 200240, China}
\newcommand{\Shanghai}{Shanghai Research Center for Quantum Sciences, Shanghai 201315, China}
\begin{document}

\title{Constraint-induced breaking and restoration of ergodicity in spin-1 PXP models}

\author{Bhaskar Mukherjee}
\email{bhaskarmukherjee1991@gmail.com}
\affiliation{\WQC}
\affiliation{\Pittsburgh}

\author{Zi Cai}
\email{zcai@sjtu.edu.cn}
\affiliation{\WQC}
\affiliation{\Key}
\affiliation{\Shanghai}

\author{W. Vincent Liu}
\email{wvliu@pitt.edu}
\affiliation{\Pittsburgh}
\affiliation{\WQC}
\affiliation{\TDLee}
\affiliation{\Shanghai}
\affiliation{\SUSTECH}

\date{\today}

\begin{abstract}
{\it Eigenstate Thermalization Hypothesis} (ETH) has played a pivotal role in understanding ergodicity and its breaking in isolated 
quantum many-body systems.
Recent experiment on 51-atom Rydberg quantum simulator and subsequent theoretical analysis have shown that hardcore kinetic
constraint can lead to weak ergodicity breaking. In this work, we demonstrate, using 1d spin-1 $PXP$ chains, that miscellaneous
type of ergodicity can be realized by adjusting the hardcore constraints between different components of nearest neighbor
spins. This includes ETH violation due to emergent shattering of Hilbert space into exponentially many subsectors 
of various sizes, a novel form of non-integrability with an extensive number of
local conserved quantities and strong ergodicity. We analyze these different forms of ergodicity and study their impact on the 
non-equilibrium dynamics of a $\mathbb{Z}_2$ initial state. We use forward scattering approximation (FSA) to understand
the amount of $\mathbb{Z}_2$-oscillation present in these models. Our work shows that not only ergodicity breaking but an
appropriate choice of constraints can lead to restoration of ergodicity as well.
\end{abstract}


\maketitle

\section{Introduction}
\label{intro}
Eigenstate thermalization hypothesis (ETH) offers the most widely accepted mechanism of thermalization of local observables 
in out of equilibrium closed quantum many body systems \cite{rev1,deustch1,srednicki1,rigol1}. An ETH satisfying system, 
prepared in an unentangled product state,
gets strongly entangled quickly under its own dynamics, losing all the information of the initial state except the conserved 
quantities (e.g total 
energy). These systems are usually strongly interacting in nature which makes the full quantum system, though well
isolated from external environment, suffer from the presence of an indigenous heat bath. This makes the study of ETH-violating
systems not only of fundamental importance but also of technological importance from the perspective of quantum information 
protection, quantum state preparation and preservation of quantum coherence (which is the measure of quantumness of a system)
up to very long time.

Integrable systems \cite{sutherland}, possessing an extensive number of conserved quantities, have long been known to disobey 
ETH. A prototypical
example is transverse field Ising model (TFIM) which though appears an interacting system in original spin language, becomes
a free system via Jordan-Wigner transformation \cite{sachdev}. Many body localized (MBL) systems \cite{mblref} which are mostly 
one-dimensional interacting 
quantum system with onsite disorder potential, forms another class of ETH violating system and has been studied in detail over 
the last decade. These systems are examples where we see strong violation of ETH in the sense all eigenstates violates ETH. 

Recently, anomalous oscillation from a density wave ($\mathbb{Z}_2$) state observed in a quench dynamics experiment 
using a 51-atom Rydberg quantum simulator \cite{expt} has revitalized the interest into the field of
thermalization and its violation. This phenomenon is understood by using spin-1/2 $PXP$ model which hosts extensive number of 
ETH violating states with high $\mathbb{Z}_2$-overlap, dubbed as quantum many body scars, in its spectrum\cite{abanin1}. 
So far, a plethora of study \cite{olexei,Shiraishi,khemani1,SU2}
has not only revealed the full phenomenology of the scar states but they have been found in a variety of models ranging from
different spin models \cite{iadecola1,moudgalya1}, Hubbard models \cite{etapairing}, higher spin PXP models\cite{TDVP,Bull},
Floquet systems \cite{Floquet}, disordered systems \cite{onsagerscar}, quantum Hall system \cite{Hall}, in higher dimension 
\cite{higherdimension}, via confinement \cite{confined} etc. In many of these studies scar states are exactly constructed 
either in matrix product state (MPS) form or by repeatedly acting suitably designed creation operator on some mother state. These kind of construction
quite straightforwardly explains the ETH violating nature of the scar states. But in spite of being the first experimentally
realized model to host quantum many body scar, a fully satisfactory explanation of scarring in $PXP$ model is still a open 
problem. Except a few scar states for which exact MPS representation was obtained \cite{olexei}, numerics and semi-analytical 
techniques like forward scattering approximation \cite{abanin1,abanin2,SU2}, single-mode approximation \cite{iadecola4} etc are 
the only option to study the majority number of scar states in PXP model.

The spin-1 $PXP$ 
model \cite{TDVP, Bull} is defined by the Hamiltonian :
$H=-\sum_i\mathcal{P}S^x_i\mathcal{P}$ in a chain of $L$ sites where the local (per site) Hilbert space 
is spanned by the eigenstates of $S^z$ ($\ket{m} \equiv \ket{-},\ket{0},\ket{+}$ for $m=-1,0,+1$) and the operator 
$\mathcal{P}=\prod_i\mathcal{P}_{i,i+1}$ characterizes the 
constrained Hilbert space. In traditional spin-1 $PXP$ \cite{TDVP} model at least one of two consecutive spin must be in the 
$\ket{-}$ state which fixes the form of the projector : 
$\mathcal{P}_{i,i+1}=P_i+P_{i+1}-P_iP_{i+1}$ with 
$P_i=\ket{-}_i\bra{-}_i$. This means, $\ket{00},\ket{+0}
,\ket{0+}$ and $\ket{++}$ type of configurations are not allowed in the constrained Hilbert space ($\mathbb{H}^{PXP}$).
This opens up the question that what happen when different set of constraints are used. In this work we show that
 the many body spectrum of spin-1 $PXP$ model can get dramatically changed when certain constraints are abolished. 
To this end we consider three different set of constraints and construct three corresponding model Hamiltonians.

{\it Three Models:} Model-I, II \& III are defined by the following Hamiltonians
\begin{equation}
 H^{\alpha}=\sum_{i=1}^L\mathcal{P}^{\alpha}S^x_i\mathcal{P}^{\alpha}
\end{equation}
where $\mathcal{P}^{\alpha}=\prod_i\mathcal{P}^{\alpha}_{i,i+1}$ for $\alpha=I,II,III$ and
$\mathcal{P}^I_{i,i+1}=\mathcal{P}_{i,i+1}+[(\ket{+}\bra{+})_i\otimes(\ket{+}\bra{+})_{i+1}]$, 
$\mathcal{P}^{II}_{i,i+1}=\mathbb{I}_i\otimes\mathbb{I}_{i+1}-[(\ket{0}\bra{0})_i\otimes(\ket{0}\bra{0})_{i+1}]$,
$\mathcal{P}^{III}_{i,i+1}=\mathcal{P}^{II}_{i,i+1}-[(\ket{+}\bra{+})_i\otimes(\ket{+}\bra{+})_{i+1}]$. 
Note that, these three models are also in $PXP$ form, but to distinguish them from traditional spin-1 $PXP$ model we use the 
model index ($\alpha$) in the superscript.

\begin{figure}
 \includegraphics[width=\linewidth]{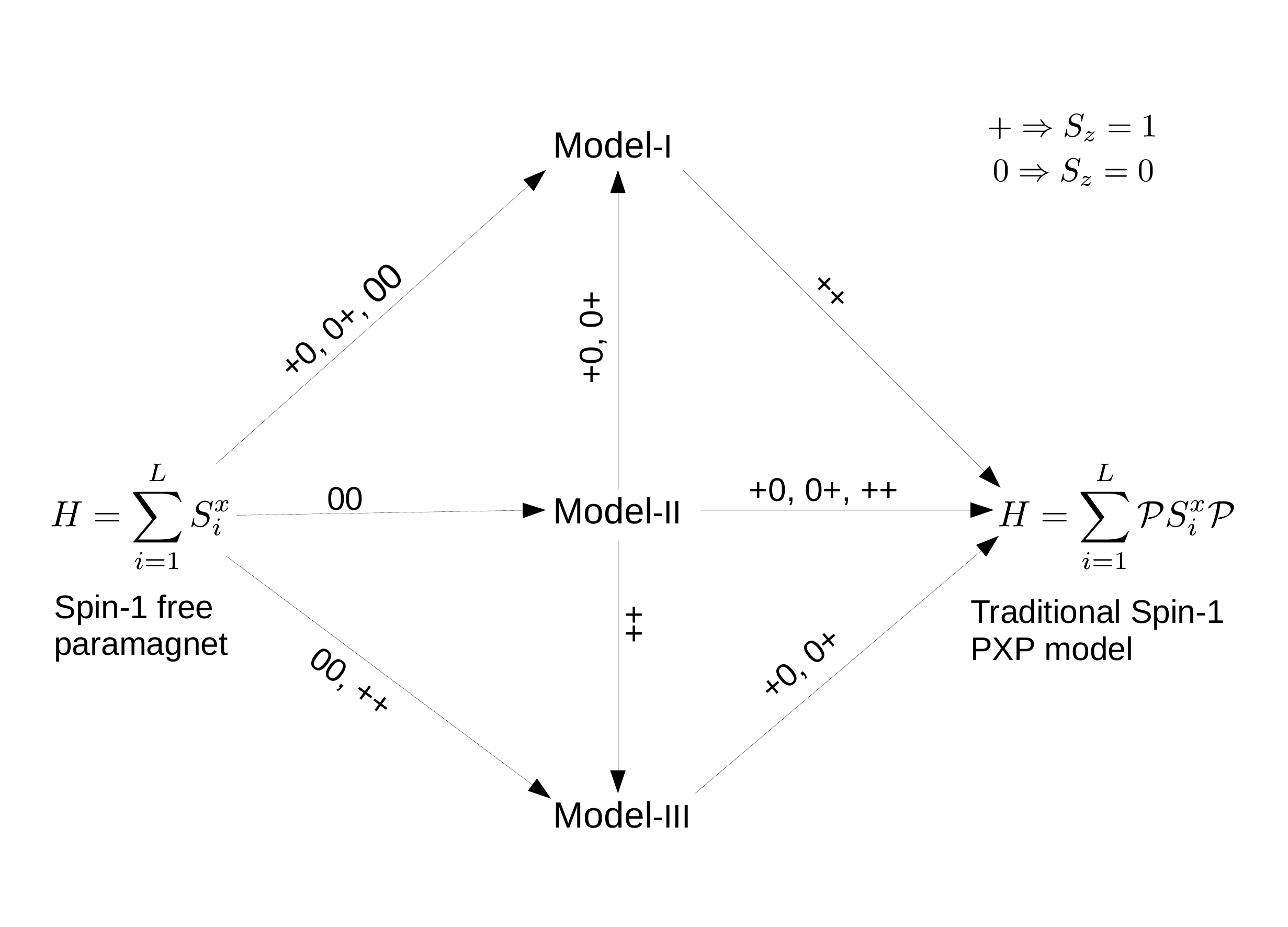}
 \caption{Schematic picture of construction of Model-I, II \& III by imposing (along the arrow direction) constraints on the 
 Hilbert space of a  spin-1 free paramagnet and abolishing (opposite to the arrow direction) constraints from traditional 
 spin-1 $PXP$ model. 
 The constraints (forbidden configurations on a pair of adjacent sites) are shown on top of the arrows.}
 \label{fig1}
\end{figure}

In Fig.\ref{fig1} we show how these models can be obtained by imposing specific constraints over a spin-1 free paramagnet
($H=\sum_iS^x_i$) or abolishing specific constraints from traditional spin-1 $PXP$ model.
We find that if we allow $\ket{++}$ configurations on top of $\mathbb{H}^{PXP}$, which we call Model-I, the 
spectrum gets shattered into exponentially many emergent subsectors of different size including $\mathbb{H}^{PXP}$
as one of the largest block. If we further allow $\ket{+0}/\ket{0+}$ type configurations (Model-II), an extensive number of 
local conserved quantities arises. This does not make Model-II exactly solvable due to some degeneracies in the spectrum
of the conserved quantities. In fact, we find that these conserved 
quantities can be used atmost to label different sectors of Model-II which are nothing but disconnected patches of spin-1/2 
$PXP$ model of 
different sizes. If we add $\ket{++}$ type configurations on top of Model-II, all the scar states disappear and the spectrum 
become strongly ergodic (Model-III). We demonstrate the consequence of these different type of ergodicity in the 
non-equilibrium dynamics of the $\mathbb{Z}_2$ state. Finally we use FSA to understand the degree of $\mathbb{Z}_2$-oscillation 
as well as ergodic nature of these three models.

The symmetries of these Hamiltonians include translation, inversion about center bond/site and particle-hole symmetry
characterized by the vanishing anticommutator of the operator $\mathcal{C}=\prod_i(2(S^z_i)^2-I_i)$ with the Hamiltonians 
: $\{H^{\alpha},\mathcal{C}\}=0$ for $\alpha=I,II,III$. The last symmetry guarantees that if there is an eigenstate $\psi$ at 
energy $E$ then there will also be an eigenstate ($\mathcal{C}\ket{\psi}$) at $-E$. 
The Hilbert space dimension grows much slowly 
than naive $3^L$ depending on the nature of constraints and choice of boundary conditions (see Table.\ref{table1}). 
We utilize the first two symmetry and work in zero momentum and inversion symmetric 
($K=0,I=+1$) sector to access largest possible system. The intertwining of the particle-hole and inversion symmetry generates
an exponentially large number of zero modes \cite{zeromode}.

\begin{figure*}[!]
 \centering
 \includegraphics[width=0.33\linewidth,height=0.26\linewidth]{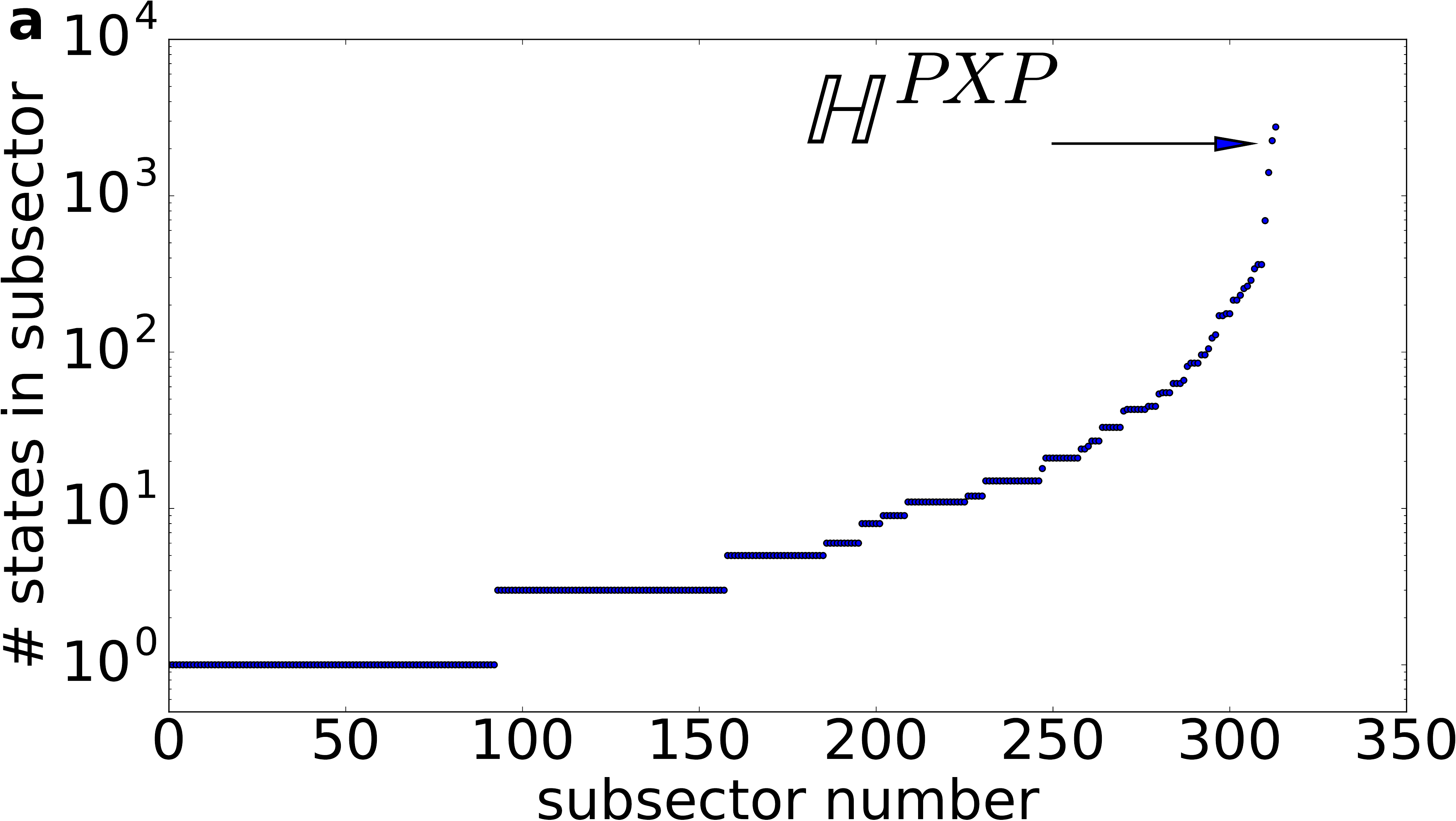}\includegraphics[width=0.33\linewidth,height=0.26\linewidth]{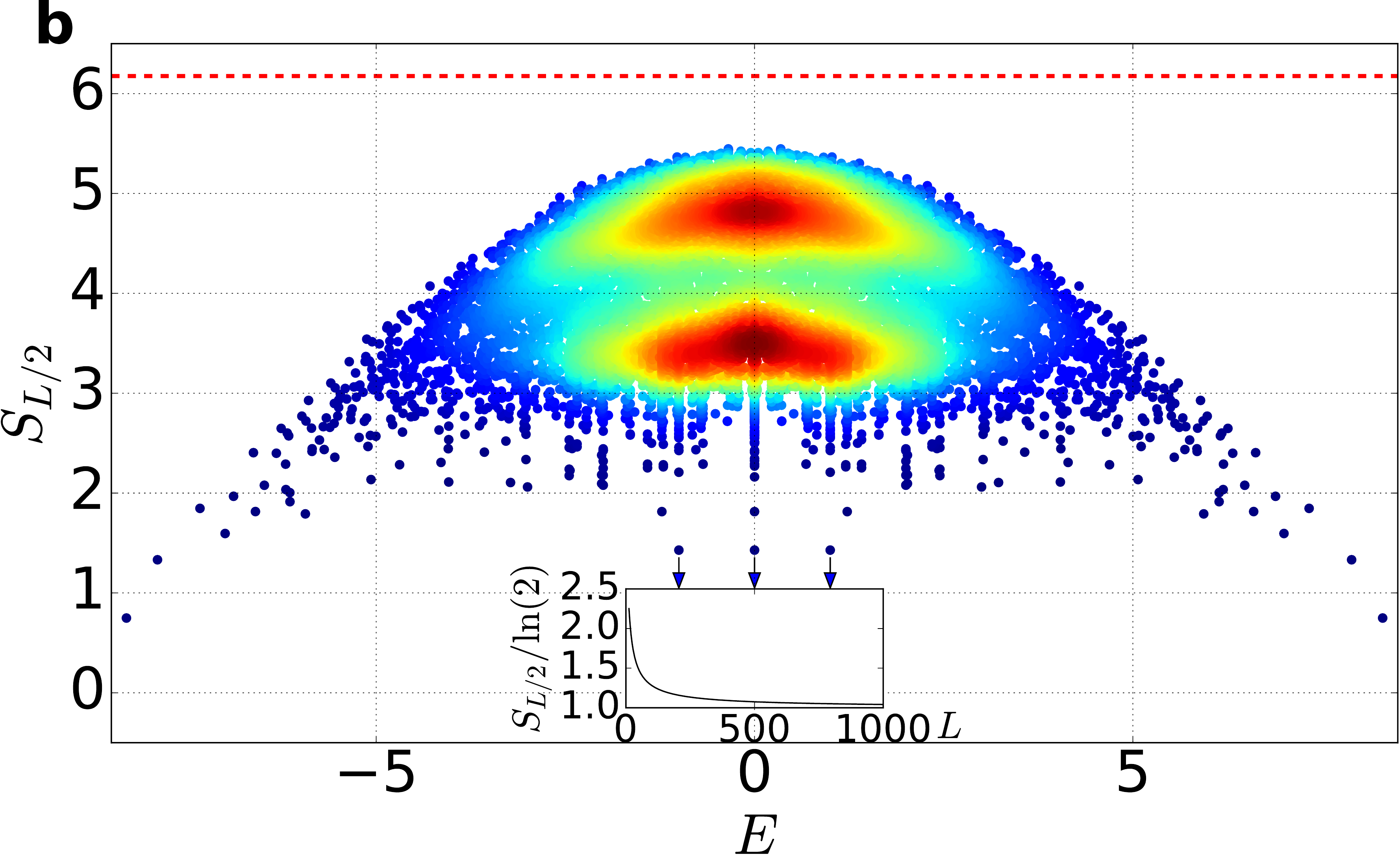}\includegraphics[width=0.33\linewidth,height=0.26\linewidth]{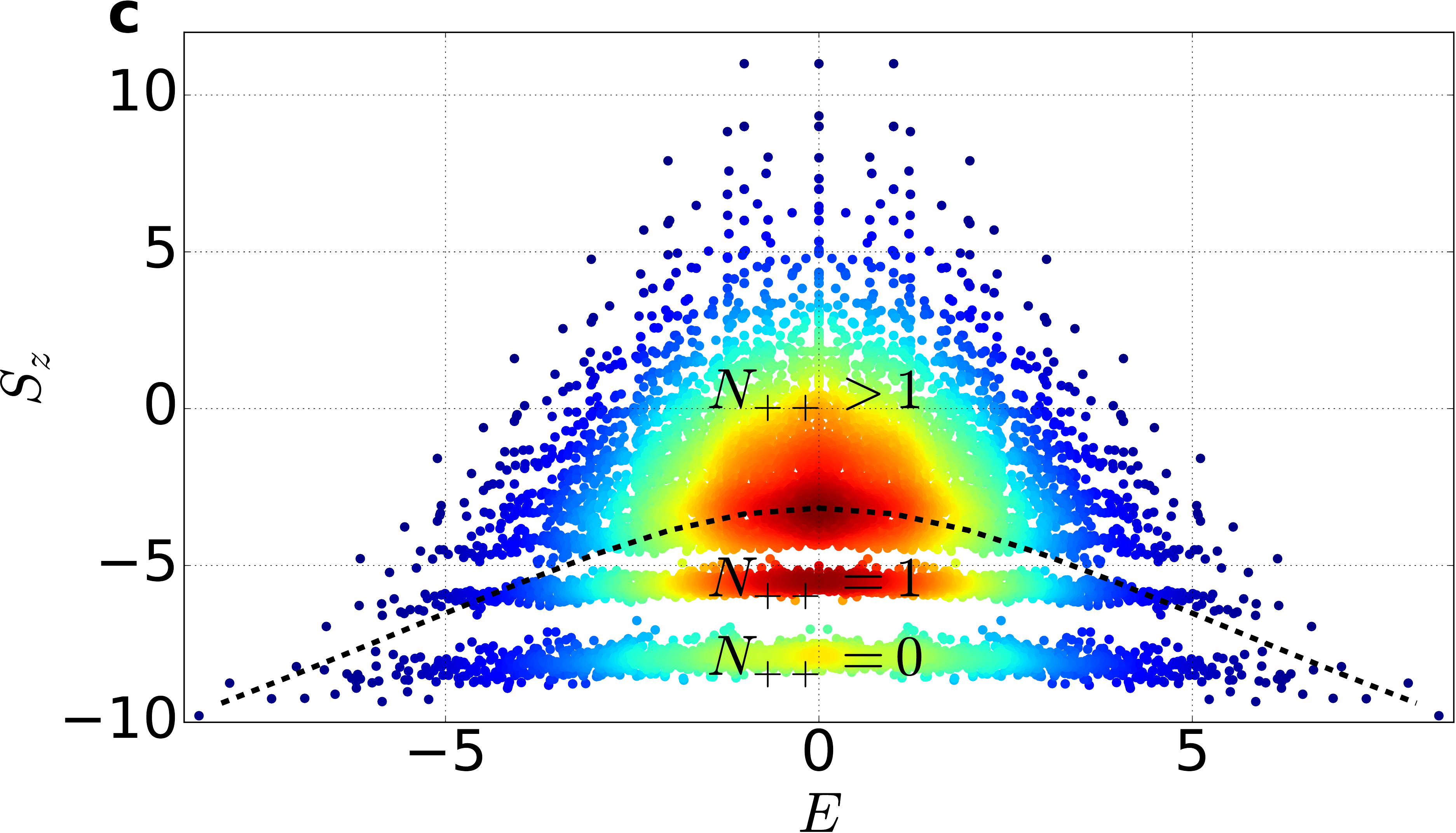}

 \includegraphics[width=0.33\linewidth,height=0.25\linewidth]{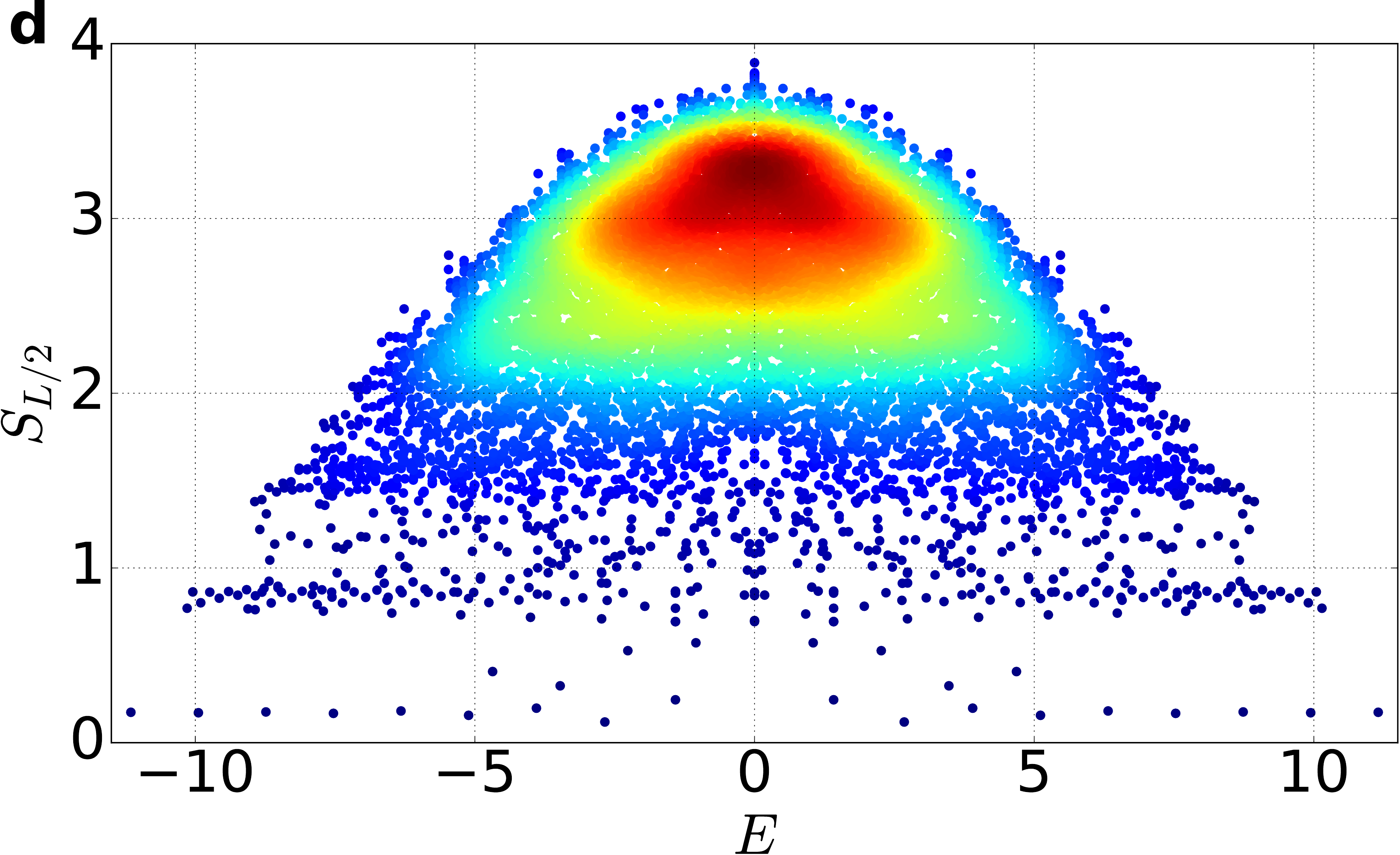}\includegraphics[width=0.33\linewidth,height=0.25\linewidth]{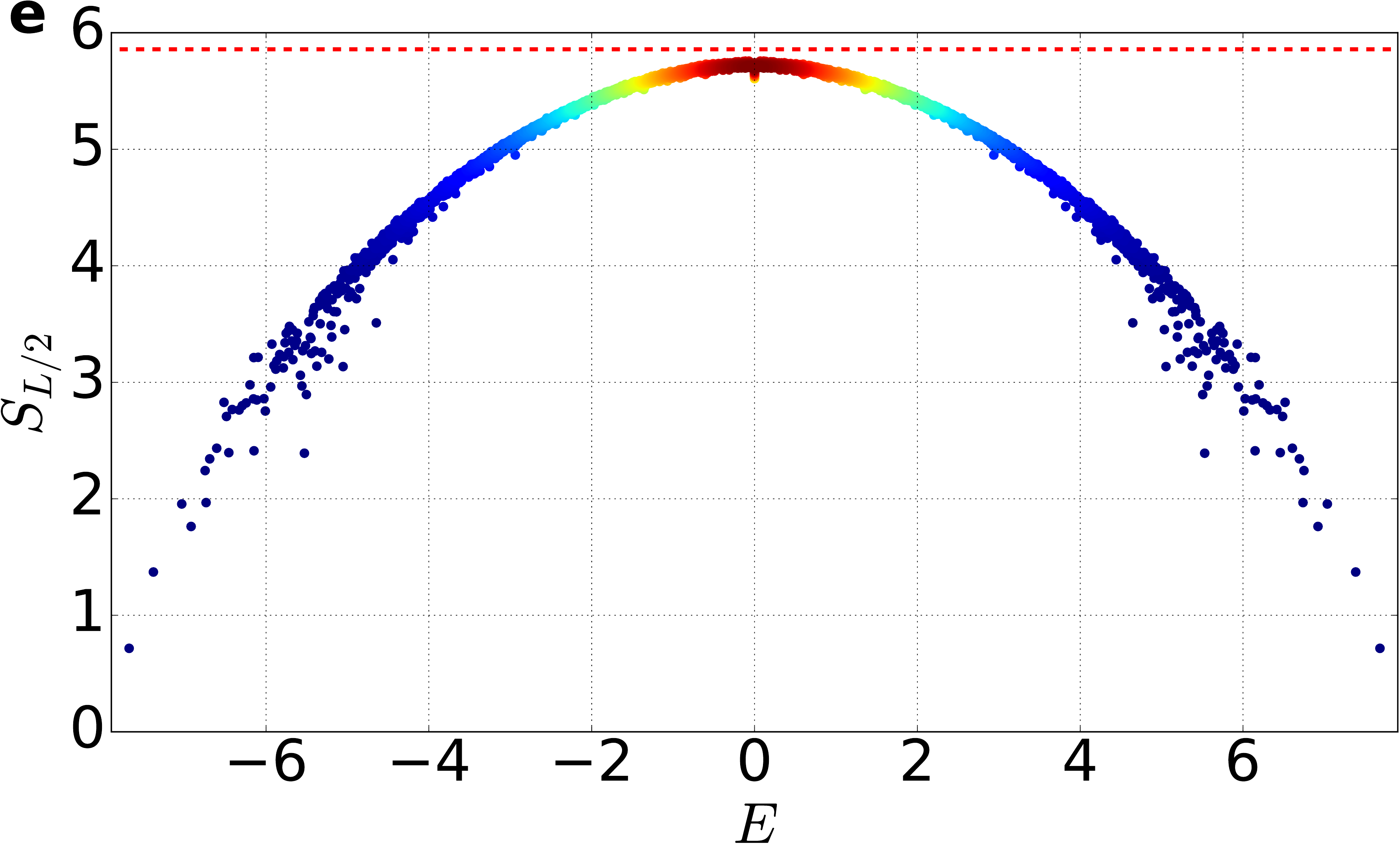}\includegraphics[width=0.33\linewidth,height=0.25\linewidth]{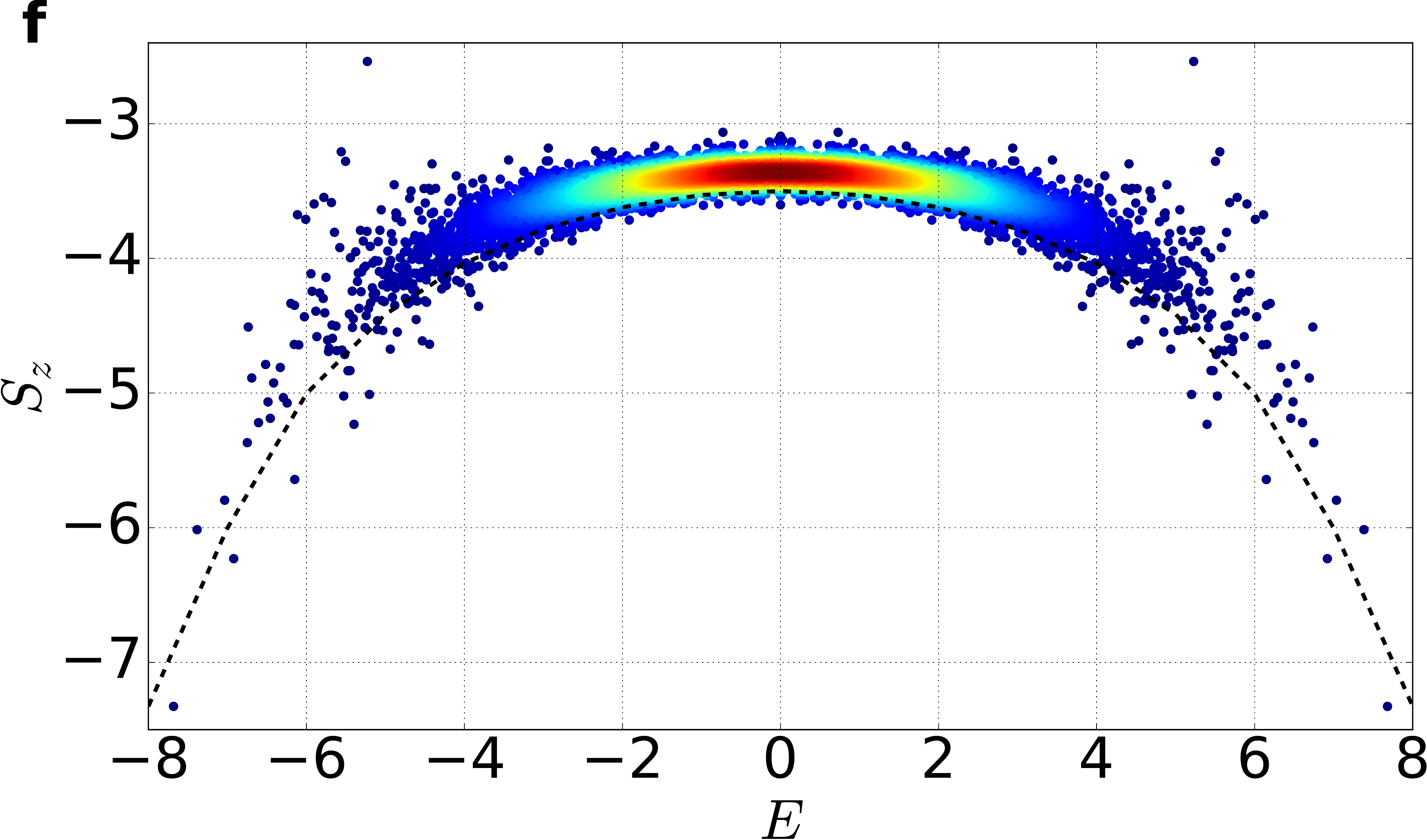}

 \caption{Ergodicity property of Model-I, II \& III. (a) Shattering of Hilbert space in Model-I. Size of different emergent
 subsectors in $K=0,I=+1$ sector for $L=16$. The traditional spin-1 $PXP$ model is the 2nd largest sector at this system size. 
 (b) $S_{L/2}$ of all the eigenstates in $K=0,I=+1$ sector of Model-I for $L=16$. System size scaling of $S_{L/2}$ for the three 
 lowest entangled state is shown in the inset. (c) $S_z$ of all eigenstates in $K=0,I=+1$ sector of Model-I for $L=16$. 
 The three cluster of states is characterized by the value of $N_{++}$. (d) $S_{L/2}$ of all the eigenstates of Model-II 
 which belong to the sectors where central $n$ ($n=2,\cdots L-2$) sites are labeled by $O=+1$ and the rest are $O=-1$, $L=20$.
 (e) $S_{L/2}$ and (f) $S_z$ of all the eigenstates in $K=0,I=+1$ sector of Model-III, $L=14$. In Panel (b),(e) the Page value
 of $S_{L/2}$ is shown in red dashed line. For panel (d) Page value ($S_{L/2}^{Page}=9.625$) is not shown. In Panel (c) and (f) 
 the canonical (Gibbs) ensemble prediction of $S_z$ is shown in black dashed line.}
 \label{fig2}
 \end{figure*}

\begin{table*}[!]
{\renewcommand{\arraystretch}{2}%
 \begin{tabular}{ |m {2cm} |m {2.5cm} |m {5.6cm}| m{4.4cm}|}
 \hline
  & Model-I & Model-II & Model-III \\ 
 \hline \hline
Forbidden configurations & $\ket{00},\ket{+0},\ket{0+}$ & $\ket{00}$ & $\ket{00},\ket{++}$\\
\hline
$d_L^{OBC}$ & $\approx2.247^L$ & 
$\frac{(1-\sqrt{3})^L(\sqrt{3}-2)+(1+\sqrt{3})^L(\sqrt{3}+2)}{2\sqrt{3}}$ $\approx2.732^L$ & 
$\frac{(1-\sqrt{2})^{L+1}+(1+\sqrt{2})^{L+1}}{2}$ $\approx2.414^L$\\
\hline
Feature & Emergent Hilbert space shattering & Exponentially many local conserved quantities ; non-integrable & 
Strongly Ergodic.\\
 \hline \hline
 \end{tabular}
\caption{Hardcore constraints, scaling of Hilbert space dimension in open boundary condition (OBC)(see \cite{supp} for derivations) and main features of 
the spectrum of Model-I, II \& III.}
\label{table1}
}

\end{table*}
The spectrum of Model-I posses a lot of degeneracies
at nonzero (including integers). We explore the connectivity
of states in the Hilbert space of Model-I and find that it (even each momentum and inversion symmetry sector) is shattered into 
exponentially many emergent subsectors \cite{khemani2,pollman1}(see Fig.\ref{fig2}(a)). The lowest possible size (in $S^z$ basis) of such emergent blocks is one, hence 
these are unentangled, zero energy eigenstates of $H^I$. These inert states remain frozen under the dynamics generated by the Hamiltonian.
The number of such inert eigenstates
($I_L$) scales as $I_L\sim \phi^L$ in large $L$ limit where 
$\phi=\frac{\sqrt{5}-1}{2}$ is the Fibonacci number and the proportionality constant depends on the choice of boundary condition
(see \cite{supp}). The eigenstates in low ($>1$) dimensional subsectors have very small but nonzero entanglement,
some of them also have interesting properties like integer energy and magnetization. The minimally entangled 
states (in the central region of the spectrum) are eigenstates of a subsector of size $3\times3$. These eigenstates (with energy 
$E=\pm1,0$) are given by $\ket{\psi_E}=\frac{1}{2}\ket{\psi_-}-\frac{e^{i\pi E}}{2}\ket{\psi_+}+\frac{E}{\sqrt{2}}\ket{\psi_0}$ 
where $\ket{\psi_m}=\frac{1}{\sqrt{L}}\sum_{n=1}^LT^n[\otimes_{i=1}^{L-3}\ket{\red{+}}_i\otimes\ket{\red{-}\blue{m}\red{-}}]$;
$m=\pm,0$ and $T$ is the translation operator. Red (blue) color is used to denote inert (active) sites. 
The half chain entanglement entropy ($S_{L/2}$) of these states are 
found to be $S_{L/2}=\ln(\frac{2L}{L-4})+\frac{4}{L}\ln(\frac{L-4}{2})$ (see \cite{supp}) which assumes area law behavior in thermodynamic limit (see
inset of Fig.\ref{fig2}(b)). We also find that some special states with $E=\pm2$ (belonging to subsectors of size $6\times6$) 
has logarithmic entanglement entropy ($S_{L/2}=\ln L-\ln2$ ; see \cite{supp}). The magnetization ($S_z=\sum_{i=1}^LS^z_i$)
of such eigenstates with energy $E=\pm n$ turns out to be $L-5n$ which are non-negative at any system size as these type of 
states don't appear for $L<5n$ (see \cite{supp}). This expectation value is much higher than the corresponding thermal 
average ($\langle S_z\rangle_{\beta}=Tr[\rho_{\beta}S_z]$ with $\rho_{\beta}=\exp(-\beta H^{I})/Tr[\exp(-\beta H^{I})]$) which
is always negative due to the excessiveness of $\ket{-}$ compared to $\ket{+}$ states in the constrained Hilbert 
space of Model-I. Thus such states clearly violates ETH. 

The number of $\ket{++}$ configurations ($N_{++}=\sum_i(\ket{++}\bra{++})_{i,i+1}$)
turns out to be a conserved quantity for Model-I which can be used to label (not uniquely) different subsectors.
The block with $N_{++}=0$ is nothing but the traditional spin-1 $PXP$ model. Note that, this kinetic constraint 
(no two spin in $\ket{+}$ state can sit next to each other) is emergent in nature as the underlying Hamiltonian does not have it \cite{iadecola3}.
The block with $N_{++}=1$ is the largest block in the 
largest possible system size ($L=16$) we have explored numerically. Blocks with same $N_{++}$ ($>1$) can be further labeled 
by $N_{+++}$ and so on. We note that there are many subsectors with $N_{++}>1$ and $N_{+++}=0$ whose states contains isolated $\ket{++}$ 
configurations, separated by active sites. Such subsectors
can be further labeled by number of $\ket{++-++}$, $\ket{++--++}$ etc type of inert configurations. For example, at $L=10$, there are 3 
subsectors with $(N_{++}=2,N_{+++}=0)$ which can be uniquely labeled by the quantum numbers ($N_{++-++},N_{++--++}$) with values
$(0,0)$, $(0,1)$ and $(1,0)$. For $L>10$, number of such subsectors is $>3$, consequently unique labeling of them can not be 
achieved. In fact, projectors on inert 
configurations of all system sizes $\leqslant L$ are conserved quantities of a system of size $L$. Though the inert states 
are unentangled, the projectors on them are nonlocal in nature. Moreover, their number scales exponentially with system size 
and thus an unique set of conserved quantities to distinguish different subsectors is lacking. That's why this is an emergent 
shattering of Hilbert space induced by the choice of constraints. We note that ETH violation due to emergent fragmentation of 
Hilbert space was first observed in fractonic circuit with local conservation of charge and dipole moment \cite{khemani2} and in
the corresponding Hamiltonian system \cite{pollman1}. 
Subsequent works showed that Hilbert space fragmentation can also happen from strict confinement\cite{iadecola2}. 

In Model-II only $\ket{00}$ type configurations are not allowed. The spectrum of Model-II also holds nonzero-energy-degeneracies 
inside the $K=0,I=+1$ sector. This time we find an extensive ($=L$) number of local quantities ($O_i$) which commutes with the 
Hamiltonian, i.e $[O_i,H^{II}]=0,\forall i$.  We find that $O_i=(\ket{+}\bra{-}+\ket{0}\bra{0}+\ket{-}\bra{+})_i=2(S^x_i)^2-\mathbb{I}_i$
(see \cite{supp}). Interestingly, this does not make Model-II completely integrable as each $O_i$ have degenerate eigenstates.
The eigenvalues of $O_i$ are $(+1,+1,-1)$ with corresponding eigenstates $\ket{O^+_1}=\frac{1}{\sqrt{2}}(\ket{+}+\ket{-}),
\ket{O^+_2}=\ket{0},\ket{O^-}=\frac{1}{\sqrt{2}}(\ket{+}-\ket{-})$. Therefore all eigenstates of $H^{II}$ can't be uniquely 
labeled by the eigenvalues of all $O_i$'s as the number of such available quantum numbers ($2^L$) is far less than the total
number of eigenstates ($\approx2.7^L$). Thus the conserved quantities can only be used to block diagonalize the Hamiltonian.
The largest block is characterized by $O_i=+1,\forall i$. It is easy to see that this sector is equivalent to
the celebrated spin-1/2 $PXP$ model with the identification $\ket{\downarrow}=\ket{O^+_1}$ and $\ket{\uparrow}=\ket{O^+_2}$.
Thus, in principle one can expect to see all phenomenon observed in spin-1/2 $PXP$ model in this spin-1 model also. On 
the theoretical side, Lin-Motrunich\cite{olexei} type exact eigenstates can also be constructed here, not only for the largest sector 
but also for all sectors where island of sites with $O_i=+1$ are separated by sites with $O_i=-1$ as these
sectors are nothing but disconnected patches of spin-1/2 $PXP$ model of different sizes. The 
smallest sector is of size 1 with $O_i=-1$ at all sites, hence $H^{II}$ posses an unentangled, exact zero energy eigenstates:
$\prod_i\ket{O^-_i}$. We note that sectors where at least two consecutive sites have $O_i=+1$ feels the 
hardcore interaction whereas sectors where each $O_i=+1$ site is isolated by at least one $O_i=-1$ from both side 
is basically non-interacting in nature. The number of such non-interacting sectors scales as $\sim\phi^L$.
Thus though many sectors of this model is non-integrable in nature there exist 
exponentially many eigenstates which violates $ETH$.

Model-III does not allow $\ket{00}$ and $\ket{++}$ type configurations. This model neither have any conserved quantity (other 
than translation, inversion and particle-hole) nor its spectrum have any emergent shattering. Due to the high connectivity 
in Hilbert space, this model displays strongly ergodic behavior (see Fig.\ref{fig2}(e),(f)). In the next section we will analyze 
it more using FSA and $\mathbb{Z}_2$-dynamics.

\begin{figure}
 \includegraphics[width=\linewidth]{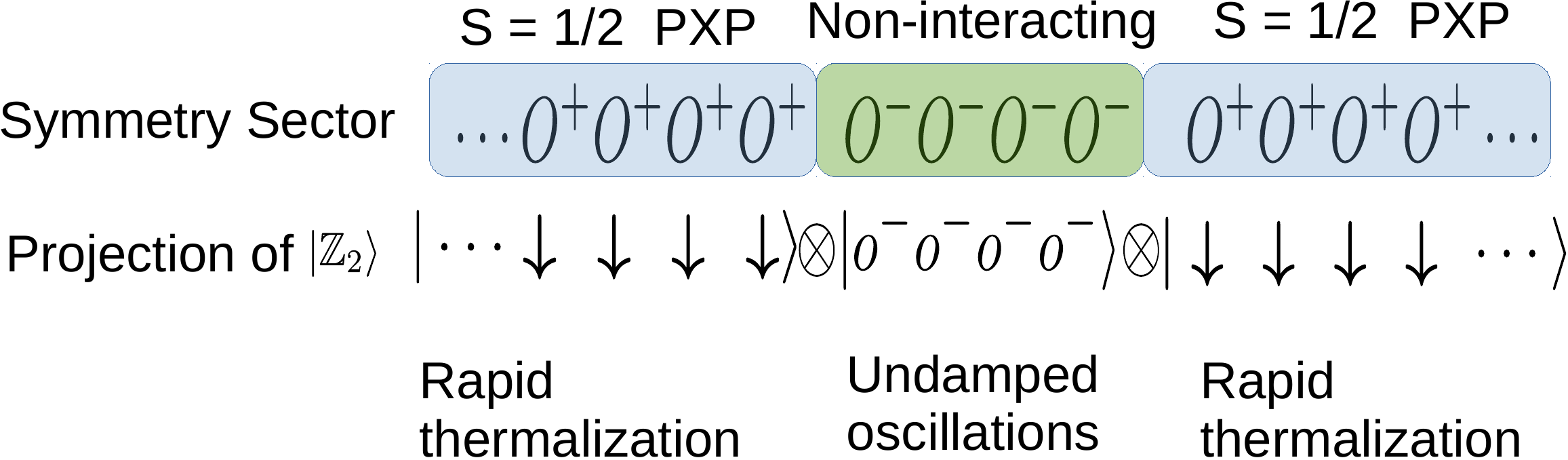}
 \caption{Schematic structure of a typical symmetry sector of Model-II and projection of $\mathbb{Z}_2$ state in that sector.}
 \label{model2schematic}
\end{figure}

\begin{figure}
 \includegraphics[width=\linewidth]{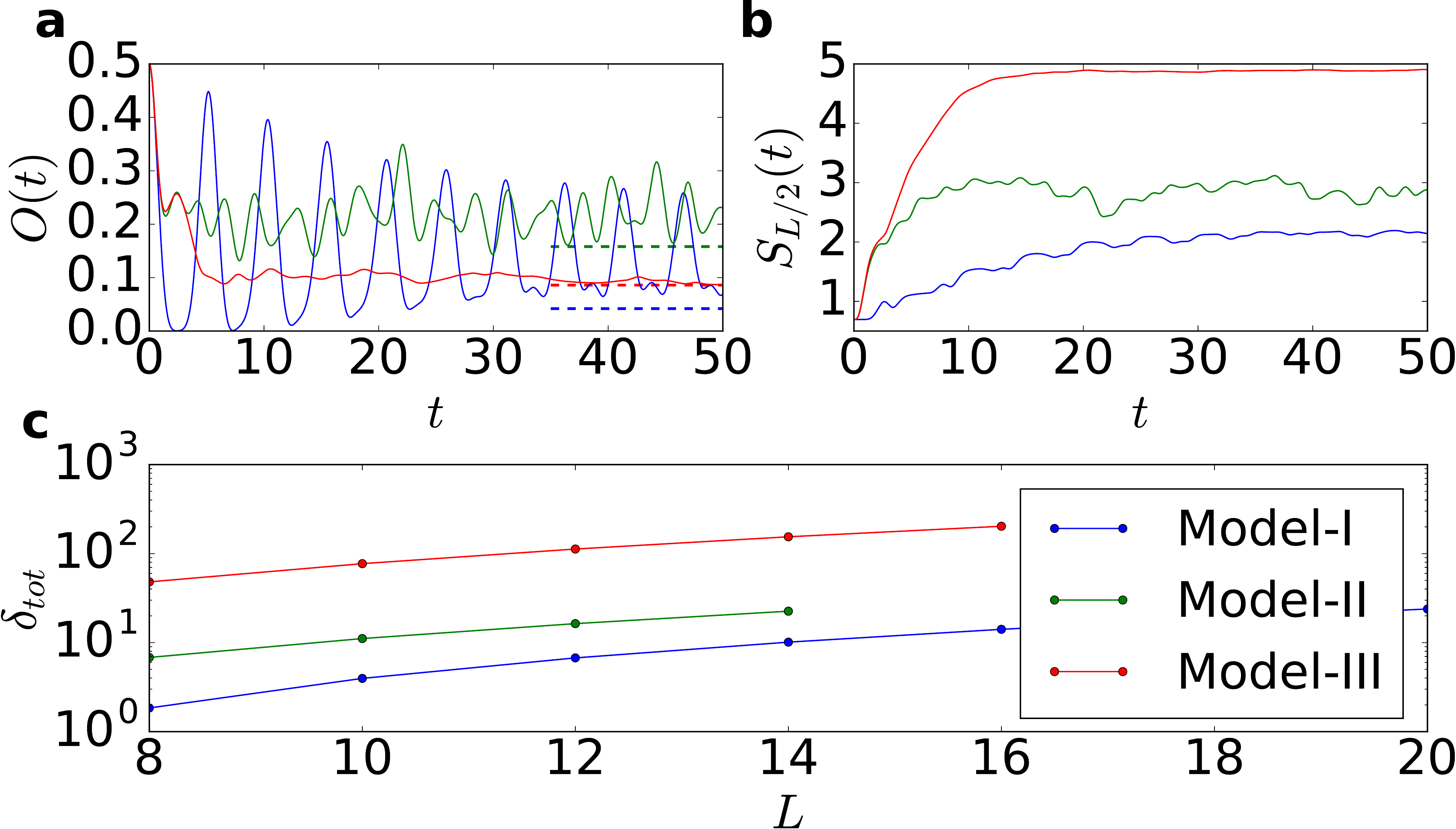}
 \caption{(a) Dynamics of a local observable (see text) in the three models. $L=12$ (b) Dynamics of $S_{L/2}$. $L=12$ (c)
 Behavior of total FSA errors as a function of system size. }\label{fig3}
\end{figure}

{\it $\mathbb{Z}_2$ dynamics and FSA :} The difference in ergodicity of Model-I, II \& III can be probed by 
studying the dynamics of local observables (we choose $O=(\ket{+}\bra{+})_i\otimes(\ket{+}\bra{+})_{i+1}$) from the initial 
state $\ket{\mathbb{Z}_2}=\otimes^{L/2}_{i=1}\ket{-}_{2i-1}\ket{+}_{2i}$. It is worthy to point out here that all standard spin-s 
$PXP$ models studied so far exhibits long lived coherent oscillation starting from the $\mathbb{Z}_2$ state\cite{TDVP}. In our Model-I
the $\mathbb{Z}_2$ state belong to the emergent subsector $\mathbb{H}^{PXP}$ and so the corresponding dynamics will be 
exactly same as that of traditional spin-1 $PXP$ model. On the other hand, $\ket{\mathbb{Z}_2}$ state has uniform overlap with all 
$2^L$ symmetry sectors of Model-II as can be seen from the expression : $\ket{\mathbb{Z}_2}=2^{-L/2}[\otimes^{L/2}_{i=1}
(\ket{O^+_1}-\ket{O^-})_{2i-1}(\ket{O^+_1}+\ket{O^-})_{2i}]$. Now how the state evolves inside a particular sector depends 
crucially on the interacting nature of that sector. The system will have perfect revival and no dephasing inside the 
fully non-interacting sectors whereas inside the interacting sectors it will thermalize rapidly
to infinite temperature (see Fig.\ref{model2schematic}). This is because the projection of the initial state in the interacting sectors corresponds to the fully
polarized down state ($\cdots \downarrow\downarrow\downarrow \cdots$) whereas the sectors themselves are nothing but 
disconnected patches of spin-1/2 $PXP$ model of different sizes. So the $\mathbb{Z}_2$-dynamics in Model-II 
is a mixture of maximally thermal and maximally non-thermal effects, as a result neither it shows strong coherent 
oscillation nor thermalizes quickly (see Fig.\ref{fig3}(a)). We note that though the number of fully non-interacting sectors 
($\sim\phi^L$) is a vanishingly small fraction ($\sim0.81^L$) of total number of sectors in thermodynamic limit, sectors with small
interacting portions also increase exponentially with system sizes. Therefore the nature of the dynamics in thermodynamic limit
is an interesting open question. Finally, we find that the $\mathbb{Z}_2$-dynamics in Model-III thermalizes rapidly to infinite 
temperature due to the strongly ergodic nature of the model. We show entanglement dynamics of the three model in Fig.\ref{fig3}
(b). The initial growth of entanglement (which is related to the speed of information propagation) is fastest in Model-III,
slowest in Model-I and intermediate in Model-II. This is consistent with the nature of the dynamics of local observable in these 
three models.

We now corroborate our dynamics results using FSA which has been very successful in capturing the scar states in large 
$PXP$ systems\cite{abanin1,abanin2}.
We begin by decomposing $H^{\alpha}$ into two parts :
$H^{\alpha}=H^{\alpha}_++H^{\alpha}_-$ such that $H^{\alpha}_-$ annihilates the initial state $\ket{v_0}=\ket{\mathbb{Z}_2}$ and 
$H^{\alpha}_+=(H^{\alpha}_-)^\dagger$. The repeated application of $H^{\alpha}_+$ on $\ket{\mathbb{Z}_2}$ generates the FSA vectors
:$\ket{v_n}=\frac{1}{\beta_n}H^{\alpha}_+\ket{v_{n-1}}$. For the spin-1 models in this paper, $n$ runs from 0 to $2L$ after which
the state gets annihilated. The FSA errors quantified by $\delta^{\alpha}_n=||H^{\alpha}_-\ket{v_n}-\beta_n\ket{v_{n-1}}||,n=1,\cdots2L$ 
are a measure of damping
force felt by the dynamical system due to many body interaction effects. The system exhibits undamped oscillations when all the $\delta_n$ are zero which can be seen 
in a free paramagnet (no constraints) or by adding suitable perturbation of appropriate strength with the constrained model
\cite{SU2}. Though, in general, FSA errors are non-zero in $PXP$ type constrained systems considered here, it could be zero 
in first few steps depending on the nature of the constraints. For example, we find that in our Model-I, first four FSA steps
are exact (i.e error free) and error arises at fifth ($n_f$) step. For Model-II and III error arises at 3rd and 2nd FSA step 
respectively. This errors cause dephasing ; higher the errors, higher is the rate of dephasing. Here we give analytical 
expression (See \cite{supp}) of first nonzero FSA error ($\delta^{\alpha}_{n_f}$) for the three models
\begin{widetext}
\begin{eqnarray}
 \delta^{I}_{n_f(=5)}&=&\frac{12(L^3-6L^2+11L-18)}{(L-1)(L-2)(L-3)(5L^4-50L^3+175L^2-250L+144)}\nonumber\\
 \delta^{II}_{n_f(=3)}&=&\frac{50(2L-9)}{(2L-5)(6L^2-45L+95)}\nonumber\\
  \delta^{III}_{n_f(=2)}&=&\frac{1}{4(4L-11)}\nonumber\\
 \label{fsaerror}
\end{eqnarray}
\end{widetext}

In brief, $\delta_{n_f}\sim L^{-(n_f-1)}$ at large $L$, from which it is easy to see that $\delta^I_{n_f}<\delta^{II}_{n_f}
<\delta^{III}_{n_f}$. We numerically calculate the FSA errors in higher steps ($n>n_f$) and plot the behavior of total 
FSA error ($\delta_{tot}=\sum_{n=1}^{2L}\delta_n$) in Fig.\ref{fig3}. One can again see that $\delta^I_{tot}<\delta^{II}_{tot}
<\delta^{III}_{tot}$. This gives a qualitative understanding of the hierarchy of entanglement growth, amount of oscillation present 
in the dynamics of local observable and ergodicity of the models. 

{\it Conclusion \& Discussion : } In summary we have studied the change in ergodic properties of spin-1 $PXP$ model for three specific choices of constraints.
We find that whereas certain set of constraints can shatter the Hilbert space into exponentially many emergent subsectors, thus
leading to violation of ETH, some other set of constraints can destroy all the anomalous states and make the spectrum strongly 
ergodic. Our choice of constraints (only between the excited states $\ket{0}$ and $\ket{+}$ in nearest neighbor (n.n) sites) is motivated by the experimental
realization of spin-1/2 $PXP$ model in Rydberg atom systems where strong repulsive interaction between n.n atoms is
turned on only when they are simultaneously in the excited (Rydberg) states\cite{expt}.
A detail study of all such set of constraints is beyond the 
scope of the current work\cite{Bull}. But we have checked that there are other set of constraints which belong to the first category (i.e
in the class of Model-I). For example, if only $\ket{+0}$ and $\ket{0+}$ type of constraints are not allowed, then also the 
constrained Hilbert space ($d_L^{OBC}\sim2.414^L$) gets shattered into many blocks of 
different sizes. In fact the special eigenstates of Model-I (with integer energies) are also the eigenstates of this Model.
Note that this model differs from Model-I by only the presence of $\ket{00}$ type of configurations which means the inert sector
or the special eigenstates of Model-I are robust against this change of constraints. 
On the experimental side, we note that non-ergodic quantum dynamics due to emergent kinetic constraint and Hilbert space 
fragmentation is recently 
observed in tilted Fermi-Hubbard model at large tilt potential\cite{expt2}. 
Secondly, we believe that the Model-II is a
minimal interacting model where non-integrability and extensive number of local conserved quantities coexist. It will be 
interesting to explore the existence of such models in other type of systems. Finally, the 
strong ergodic nature of Model-III is also remarkable because we find that quantum scars exist and ETH violation happens even 
when only $\ket{++}$ type of configurations are forbidden. We leave the study of detailed mechanism behind this constraint induced 
strong ergodicity and the exploration of class of constraints which leads to the same behavior as a future problem.

\begin{acknowledgements}
 {\it Acknowledgments : }
This work is supported by the Shanghai Municipal Science and Technology Major Project
(Grant No. 2019SHZDZX01) [B.M., Z.C., W.V.L.], National Key Research and Development
Program of China (Grants No. 2016YFA0302001 and 2020YFA0309000) and NSFC of China (Grants No.
11674221 and No. 11574200) [Z.C] and by the AFOSR Grant No. FA9550-16-1-0006 and
the MURI-ARO Grant No. W911NF17-1-0323 through UCSB [W.V. L.].  
\end{acknowledgements}

\begin{widetext}
\appendix
\section{Supplemental material for \textquotedblleft Constraint-induced breaking and restoration of ergodicity in spin-1 PXP 
models\textquotedblright}
\section{Growth of Hilbert space dimension for Model-I, II \& III}
\label{appA}

The constraints leads to a slower growth of Hilbert space dimension compared to naive $3^L$. Atfirst we demonstrate it for open 
boundary condition (OBC). 
We start by Model-I for which $\ket{+0},\ket{0+},\ket{00}$ type of configurations are not allowed. 
Any state in a system of size $L$ may end 
by $\ket{-},\ket{0}$ or $\ket{+}$. All states in a $L$ site system which are ending by $\ket{-}$ can be obtained by simply 
appending a $\ket{-}$ to all states in a $L-1$ site system.
The states which are ending 
by $\ket{0}$ in a 
$L$ site system can be obtained by appending $\ket{-0}$ to all states in a $L-2$ site system whereas $\ket{+}$ can only be 
appended if the last site is not in the state $\ket{0}$. This leads to the following recurrence relation of 
total number of states ($d_L$) in a system of size $L$
\begin{eqnarray}
d_L&=&d_L^-+d_L^0+d_L^+\nonumber\\
&=&d_{L-1}+d_{L-2}+(d_{L-1}-d_{L-3})\nonumber\\
&=&2d_{L-1}+d_{L-2}-d_{L-3}
\end{eqnarray}
which can be cast in the following matrix form
\begin{equation}
 \left(\begin{array}{c}
        d_L\\
        d_{L-1}\\
        d_{L-2}\\
       \end{array}\right)=
       \left(\begin{array}{ccc}
              2 & 1 & -1\\
              1 & 0 &  0\\
              0 & 1 &  0\\
             \end{array}\right)\left(\begin{array}{c}
                                     d_{L-1}\\
                                     d_{L-2}\\
                                     d_{L-3}\\
                                     \end{array}\right)
\end{equation}
The eigenvalues of this matrix are $\approx(2.247, -0.802, 0.555)$. It is difficult to get an exact expression of $d_L$ for 
Model-I but the leading behavior in the large 
$L$ limit will be controlled by the largest eigenvalue 2.247 (as this is the only one with magnitude $>1$).

Model-II does not allow $\ket{00}$ type configurations only. So, $\ket{0}$ can only be appended if the last site is either in 
state $\ket{+}$ or $\ket{-}$. The recurrence relation for $d_L$ is given by
\begin{eqnarray}
d_L&=&d_L^-+d_L^0+d_L^+\nonumber\\
&=&d_{L-1}+(d^+_{L-1}+d^-_{L-1})+d_{L-1}\nonumber\\
&=&2d_{L-1}+2d_{L-2}
\end{eqnarray}
which can be cast in the following matrix form
\begin{equation}
 \left(\begin{array}{c}
        d_L\\
        d_{L-1}\\
       \end{array}\right)=
       \left(\begin{array}{cc}
              2 & 2 \\
              1 & 0 \\
             \end{array}\right)\left(\begin{array}{c}
                                     d_{L-1}\\
                                     d_{L-2}\\
                                     \end{array}\right)=
                                            \left(\begin{array}{cc}
              2 & 2 \\
              1 & 0 \\
             \end{array}\right)^{L-2}\left(\begin{array}{c}
                                     d_2\\
                                     d_1\\
                                     \end{array}\right)
\end{equation}
using $d_1=3$ and $d_2=8$ one get the expression of $d_L$ as a function of $L$ (see Table.-\ref{table1} in main text).

In Model-III 
$\ket{+0}/\ket{0+}$ type of configurations are not allowed. States which are ending by $\ket{0}(\ket{+})$ in a $L$ site system
can be obtained by appending $\ket{0}(\ket{+})$ to all states of a $L-1$ site system which are not ending by $\ket{+}(\ket{0})$.
Therefore,
\begin{eqnarray}
 d_L&=&d_L^-+d_L^0+d_L^+\nonumber\\
  &=&d_{L-1}+(d_{L-1}-d_{L-1}^+)+(d_{L-1}-d_{L-1}^0)\nonumber\\
  &=&3d_{L-1}-(d^0_{L-1}+d_{L-1}^+)\nonumber\\
  &=&2d_{L-1}+d_{L-2}
\end{eqnarray}
this recurrence relation can be represented in the following matrix form
\begin{equation}
 \left(\begin{array}{c}
        d_L\\
        d_{L-1}\\
       \end{array}\right)=
       \left(\begin{array}{cc}
              2 & 1 \\
              1 & 0 \\
             \end{array}\right)\left(\begin{array}{c}
                                     d_{L-1}\\
                                     d_{L-2}\\
                                     \end{array}\right)=
                                            \left(\begin{array}{cc}
              2 & 1 \\
              1 & 0 \\
             \end{array}\right)^{L-2}\left(\begin{array}{c}
                                     d_2\\
                                     d_1\\
                                     \end{array}\right)
\end{equation}
the exact expression of $d_L$ for Model-I can be calculated using $d_1=3$ and $d_2=7$ (see Table-\ref{table1} in main text).

Hilbert space dimension in $PBC$ will be somewhat smaller than the corresponding number in $OBC$ as some of the configurations(
which does not satisfies the constraints between the two end spins) will be eliminated. 

\section{Special states of Model-I and their properties}
\label{appB}

\subsection{Inert states}
There are exponentially many states inside the constrained Hilbert 
space of Model-I that are not connected by the Hamiltonian with other states and hence inert \cite{khemani1,pollman1}. 
Inert states are unentangled, zero-energy eigenstates of $H^{I}$. We first 
consider the most obvious one : $\ket{+++\cdots}$, the product of $\ket{S_z=1}$ state at all sites. Starting from this state, 
one can construct other inert states by inserting one or more
$\ket{-}$ in the sea of $\ket{+}$ states. There can not be three consecutive $\ket{-}$ state and isolated (i.e surrounded by 
at least one $\ket{-}$ from both side) $\ket{+}$ state, as the presence of such configurations makes the state active.
Note that, due to the same reason, presence of any $\ket{0}$ state is also not allowed. We find that the exact number of 
inert states ($I_L$) for system size $L$ follows the recurrence relation $I_L=I_{L-1}+I_{L-3}+I_{L-4}$ which can be cast 
in the following matrix form
\begin{equation}
  \left(\begin{array}{c}
        I_L\\
        I_{L-1}\\
        I_{L-2}\\
        I_{L-3}
       \end{array}\right)=
       \left(\begin{array}{cccc}
              1 & 0 & 1 & 1\\
              1 & 0 & 0 & 0\\
              0 & 1 & 0 & 0\\
              0 & 0 & 1 & 0\\
             \end{array}\right)\left(\begin{array}{c}
                                     I_{L-1}\\
                                     I_{L-2}\\
                                     I_{L-3}\\
                                     I_{L-4}
                                     \end{array}\right)
\end{equation}
Interestingly this holds for both $PBC$ and $OBC$. Therefore, raising this matrix to appropriate power and using suitable 
boundary conditions we get 
\begin{eqnarray}
I_L^{OBC}&=&\frac{1}{10}((3+\sqrt{5})\phi^L-(3-\sqrt{5})(\frac{-1}{\phi})^L+4(\cos(\frac{\pi L}{2})-2\sin(\frac{\pi L}{2})))\nonumber\\
I_L^{PBC}&=&2\cos(\frac{\pi L}{2})+\phi^L+\phi^{-L}
\end{eqnarray}
where $\phi(=\frac{1+\sqrt{5}}{2})$ is the Fibonacci number. It is easy to see that for large systems $I_L\sim c\phi^L$ where 
$c=1(0.523)$ for $PBC(OBC)$ which means there is nearly half amount of inert states in $OBC$ compared to $PBC$.
\subsection{Special states with integer energies}
Special states with energy $E=\pm1$ are given by
\begin{equation}
 \ket{\psi_{\pm1}}=\frac{1}{2}\ket{\psi_-}+\frac{1}{2}\ket{\psi_+}\pm\frac{1}{\sqrt{2}}\ket{\psi_0}
 \label{E2}
\end{equation}
where
\begin{eqnarray}
 \ket{\psi_-}&=&\frac{1}{\sqrt{L}}\sum_{n=1}^LT^n\ket{\underset{L-3}{\underbrace{\red ++\cdots++}}\red-\blue-\red-}\nonumber\\
 \ket{\psi_+}&=&\frac{1}{\sqrt{L}}\sum_{n=1}^LT^n\ket{\underset{L-3}{\underbrace{\red ++\cdots++}}\red-\blue+\red-}\nonumber\\
 \ket{\psi_0}&=&\frac{1}{\sqrt{L}}\sum_{n=1}^LT^n\ket{\underset{L-3}{\underbrace{\red ++\cdots++}}\red-\blue0 \ \red-}
 \label{E22}
\end{eqnarray}

Next, eigenstates with energy $E=\pm2$ are given by
\begin{equation}
 \ket{\psi_{\pm2}}=\pm\frac{1}{4}\ket{\psi_{--}}\pm\frac{1}{\sqrt{8}}\ket{\psi_{+-}}\pm\frac{1}{4}\ket{\psi_{++}}+
 \frac{1}{2}\ket{\psi_{0-}}+\frac{1}{2}
 \ket{\psi_{+0}}\pm\frac{1}{2}\ket{\psi_{00}}
 \label{E4}
\end{equation}

where 
\begin{eqnarray}
 \ket{\psi_{--}}&=&\sqrt{\frac{2}{L}}\sum_{n=1}^{L/2}T^n\ket{\underset{\frac{L}{2}-3}{\underbrace{{\red ++\cdots++}}}{\red-}{\blue-}{\red-}
 \underset{\frac{L}{2}-3}{\underbrace{\red ++\cdots++}}\red-\blue-\red-}\nonumber\\
 \ket{\psi_{+-}}&=&\frac{1}{\sqrt{L}}\sum_{n=1}^{L}T^n\ket{\underset{\frac{L}{2}-3}{\underbrace{{\red ++\cdots++}}}{\red-}{\blue+}{\red-}
 \underset{\frac{L}{2}-3}{\underbrace{\red ++\cdots++}}\red-\blue-\red-}\nonumber\\
 \ket{\psi_{++}}&=&\sqrt{\frac{2}{L}}\sum_{n=1}^{L/2}T^n\ket{\underset{\frac{L}{2}-3}{\underbrace{{\red ++\cdots++}}}{\red-}{\blue+}{\red-}
 \underset{\frac{L}{2}-3}{\underbrace{\red ++\cdots++}}\red-\blue+\red-}\nonumber\\
 \ket{\psi_{0-}}&=&\frac{1}{\sqrt{L}}\sum_{n=1}^{L}T^n\ket{\underset{\frac{L}{2}-3}{\underbrace{{\red ++\cdots++}}}{\red-}{\blue0}{\red-}
 \underset{\frac{L}{2}-3}{\underbrace{\red ++\cdots++}}\red-\blue-\red-}\nonumber\\
 \ket{\psi_{+0}}&=&\frac{1}{\sqrt{L}}\sum_{n=1}^{L}T^n\ket{\underset{\frac{L}{2}-3}{\underbrace{{\red ++\cdots++}}}{\red-}{\blue+}{\red-}
 \underset{\frac{L}{2}-3}{\underbrace{\red ++\cdots++}}\red-\blue0\red-}\nonumber\\
 \ket{\psi_{00}}&=&\sqrt{\frac{2}{L}}\sum_{n=1}^{L/2}T^n\ket{\underset{\frac{L}{2}-3}{\underbrace{{\red ++\cdots++}}}{\red-}{\blue0}{\red-}
 \underset{\frac{L}{2}-3}{\underbrace{\red ++\cdots++}}\red-\blue0\red-}\nonumber\\
 \label{E42}
\end{eqnarray}

Red (blue) colored sites are inert (active).
Note that there is a structural resemblance between the states in Eq.\eqref{E2},\eqref{E22}
and in Eq.\eqref{E4},\eqref{E42}. The later type of states (in a system of size $L$) is obtained by a spatial addition of 
different combinations of the former type of states (in a system of size $L/2$). This also explains the additivity of their
energies.
\subsubsection{Proof of eigenstates}
Here we prove that the special states in Eq. \eqref{E2}, \eqref{E4} are exact eigenstates of Model-I.
We find that the following relations hold for the states in Eq. \eqref{E22} and Model-I.
\begin{eqnarray}
 H^I\ket{\psi_-}&=&\frac{1}{\sqrt{2}}\ket{\psi_0}\nonumber\\
 H^I\ket{\psi_+}&=&\frac{1}{\sqrt{2}}\ket{\psi_0}\nonumber\\
 H^I\ket{\psi_0}&=&\frac{1}{\sqrt{2}}(\ket{\psi_+}+\ket{\psi_-})
\end{eqnarray}
These relations together gives $H^I\ket{\psi_{\pm1}}=\pm\ket{\psi_{\pm1}}$

Similarly for the states in Eq. \eqref{E42}, we find
\begin{eqnarray}
H^I\ket{\psi_{--}}&=&\ket{\psi_{0-}}\nonumber\\
H^I\ket{\psi_{+-}}&=&\frac{1}{\sqrt{2}}(\ket{\psi_{0-}}+\ket{\psi_{+0}})\nonumber\\
H^I\ket{\psi_{++}}&=&\ket{\psi_{+0}}\nonumber\\
H^I\ket{\psi_{0-}}&=&\ket{\psi_{--}}+\ket{\psi_{00}}+\frac{1}{\sqrt{2}}\ket{\psi_{+-}}\nonumber\\
H^I\ket{\psi_{+0}}&=&\frac{1}{\sqrt{2}}\ket{\psi_{+-}}+\ket{\psi_{++}}+\ket{\psi_{00}}\nonumber\\
H^I\ket{\psi_{00}}&=&\ket{\psi_{0-}}+\ket{\psi_{+0}}\nonumber\\
\end{eqnarray}
These relations together gives $H^I\ket{\psi_{\pm2}}=\pm2\ket{\psi_{\pm2}}$.
\subsubsection{Magnetization}
Here we show that the special eigenstates have integer magnetization ($S_z=\sum_iS^z_i$).

For states in Eq.\eqref{E2}
\begin{equation}
 \bra{\psi_{\pm1}}S_z\ket{\psi_{\pm1}}=\frac{L-6}{4}+\frac{L-4}{4}+\frac{L-5}{2}=L-5
\end{equation}
For states in Eq.\eqref{E4}
\begin{equation}
\bra{\psi_{\pm2}}S_z\ket{\psi_{\pm2}}=\frac{L-12}{16}+\frac{L-10}{8}+\frac{L-8}{16}+\frac{L-11}{4}+\frac{L-9}{4}+\frac{L-10}{4}
=L-10
\end{equation}
Similarly for states with energy $E=\pm n$, magnetization will be $L-5n$. We note that there are lots of eigenstates at integer 
energies in Model-I. The specialty of integer energy eigenstates (e.g in \eqref{E2},\eqref{E4}) studied in this work is that 
they have minimum entanglement entropy and maximum magnetization in the corresponding manifold of states. 

\subsubsection{Entanglement entropies}
The entanglement of a state can be quantified using various schemes. We use Von-Neumann formula which works in the following way :
 first divide the full system $AB$ into two parts, $A$ and $B$. Then the entropy of entanglement of part $A$ with part $B$ is 
 given by $S_A=Tr[\rho_A\ln(\rho_A)]$ where $\rho_A=Tr_B[\rho_{AB}]$ is the reduced density matrix of part $A$ and $\rho_{AB}$
 is the density matrix corresponding to the state of the full system. The size of $A$ and $B$ can be anywhere in between 1 and 
 $L-1$ with the constraint $L_A+L_B=1$. As we use $PBC$ for all our calculation, the full system $AB$ will be in $PBC$ we have
 to treat the individual part $A$ and $B$ using $OBC$. One needs to be careful while taking the partial trace of the full 
 density matrix (to ensure $Tr[\rho_A]=1$) as due to the constraints in Hilbert space, the state of $AB$ is not tensor product 
 of states in $A$ and $B$ (see Fig.\ref{fig:ent}).
 
We first derive the single site reduced density matrices ($\rho_1$) for the special states in Eq.\eqref{E2}. 
The local HSD per site is 3, so $\rho_1$ is a $3\times3$ matrix, whose diagonal elements for the states in Eq.\eqref{E2} are
\begin{eqnarray}
 \rho_1^{++}&=&\frac{1}{4}.\frac{1}{L}.(L-3)+\frac{1}{4}.\frac{1}{L}(L-2)+\frac{1}{2}.\frac{1}{L}.(L-3)\nonumber\\
 &=&1-\frac{11}{4L}\nonumber\\
 \rho_1^{00}&=&\frac{1}{2}.\frac{1}{L}\nonumber\\
 \rho_1^{--}&=&\frac{1}{4}.\frac{1}{L}.3+\frac{1}{4}.\frac{1}{L}.2+\frac{1}{2}.\frac{1}{L}.2=\frac{9}{4L}
\end{eqnarray}
The off-diagonal elements are 
\begin{eqnarray}
\rho_1^{+0}&=&\rho_1^{0+}=\frac{1}{2}.\frac{1}{\sqrt{2}}.\frac{1}{L}\nonumber\\
\rho_1^{+-}&=&\rho_1^{-+}=\frac{1}{2}.\frac{1}{2}.\frac{1}{L}\nonumber\\
\rho_1^{0-}&=&\rho_1^{-0}=\frac{1}{2}.\frac{1}{\sqrt{2}}.\frac{1}{L}
\end{eqnarray}
note that all matrix elements except one diagonal element die out in the thermodynamic limit which means 
$\displaystyle{\lim_{L \to \infty}S_1=0}$. Therefore, in the thermodynamic limit, any site is unentangled with the rest of the 
system. This holds for the state in Eq.\eqref{E4} also.

Now we concentrate on half chain entanglement entropy. Though the dimension of the corresponding reduced density matrices 
($\rho_{L/2}$) scales 
as $\approx2.249^{L/2}$, for the 
states in Eq.\eqref{E2} we find only six eigenvalues are nonzero at any system size $L$, which are $\frac{L-4}{2L}$ (multiplicity 2)
and $\frac{1}{L}$ (multiplicity 4). This gives $S_{L/2}=-2.\frac{L-4}{2L}\ln(\frac{L-4}{2L})-4.\frac{1}{L}\ln(\frac{1}{L})$ 
which
has been plotted in Fig. \ref{fig:ent}. The decrease of $S_{L/2}$ with $L$ and its saturation to the area law value ($\ln2$)
in asymptotically large system size is an artifact of the non-tensorproduct structure of the constrained Hilbert space.
For the states in Eq.\eqref{E4} we find the number of nonzero 
eigenvalues to be $L/2$ which are all equal with magnitude $\frac{2}{L}$. This gives 
$S_{L/2}=\frac{L}{2}.\frac{2}{L}.\ln(\frac{2}{L})=\ln(L)-\ln(2)$.
\begin{figure}
 \includegraphics[width=0.5\linewidth]{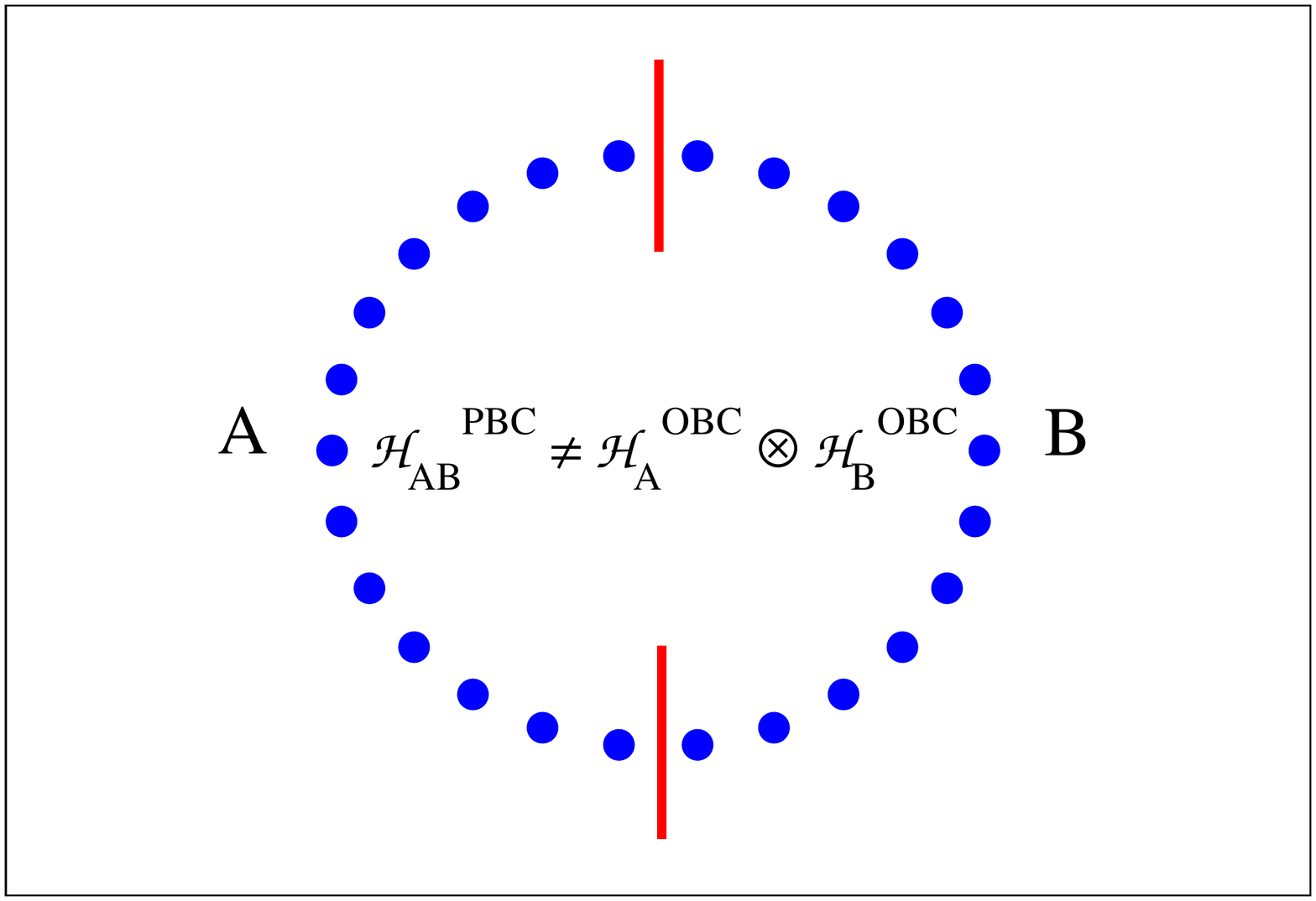}\includegraphics[width=0.5\linewidth]{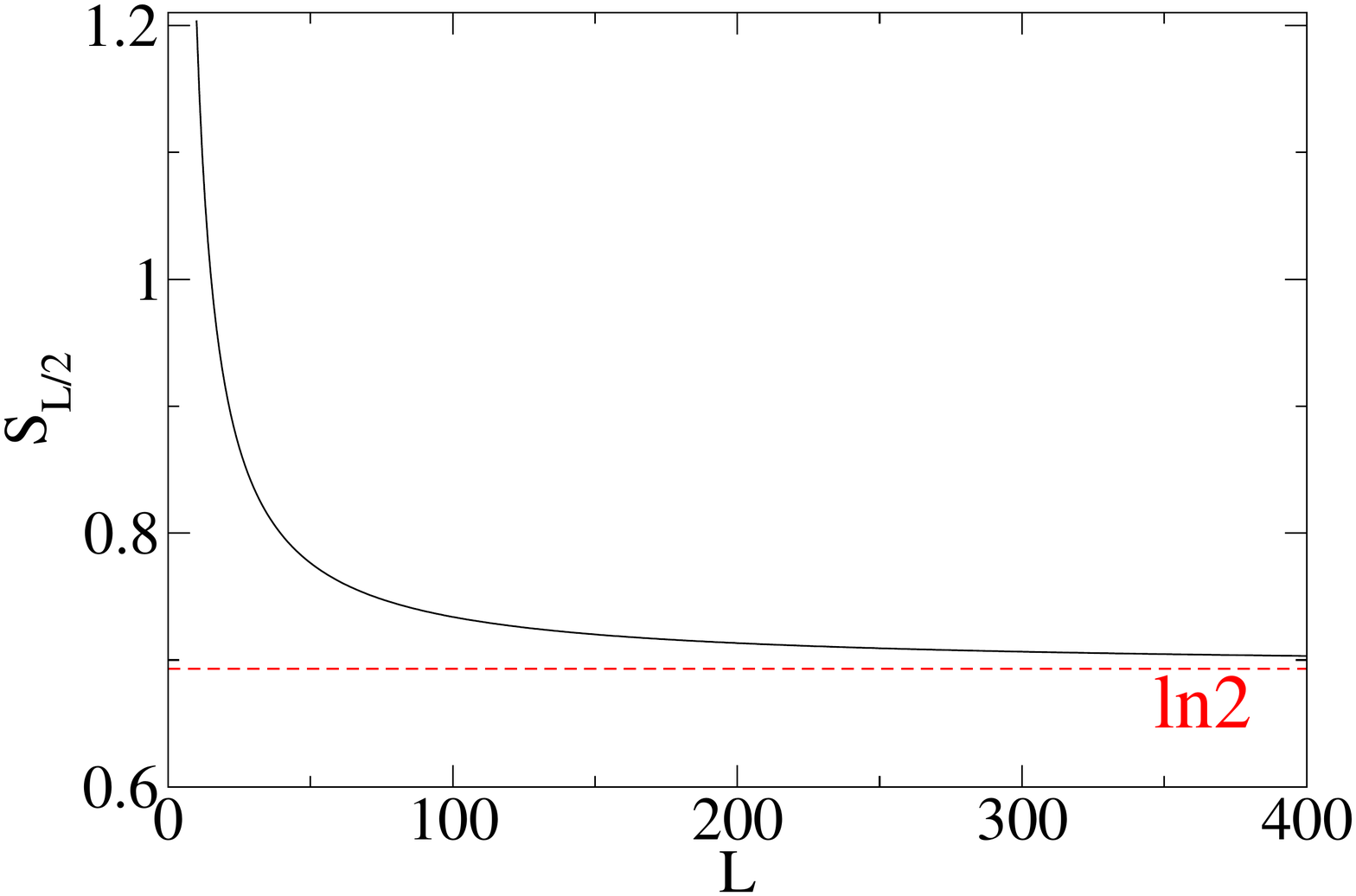}
 \caption{left : Partition of a constrained system into two parts. right : $S_{L/2}$ as a function of $L$ for the states in 
 \eqref{E2}}
 \label{fig:ent}
\end{figure}
\section{Conserved quantities of Model-II}
\label{appC}
Here we show that the operators $O_i$ commutes with the Hamiltonian $H^{II}$. We expand the relevant
portion of $H^{II}$
\begin{eqnarray}
 &&\mathcal{P}^{II}_{i-1,i}S^x_i\mathcal{P}^{II}_{i,i+1}\nonumber\\
 &=&\mathbb{I}_{i-1}\otimes S^x_i\otimes\mathbb{I}_{i+1}-\frac{1}{\sqrt{2}}\mathbb{I}_{i-1}\otimes(\ket{+}\bra{0}+
 \ket{-}\bra{0})_i\otimes
 (\ket{0}\bra{0})_{i+1}-\frac{1}{\sqrt{2}}(\ket{0}\bra{0})_{i-1}\otimes
 (\ket{0}\bra{+}+\ket{0}\bra{-})_i\otimes\mathbb{I}_{i+1}\nonumber\\
 \label{H3local}
\end{eqnarray}
where we have used that $S^x_i=\frac{1}{\sqrt{2}}(\ket{+}\bra{0}+\ket{0}\bra{+}+\ket{0}\bra{-}+\ket{-}\bra{0})_i$ and 
$\mathcal{P}^{II}_{i,i+1}=\mathbb{I}_i\otimes\mathbb{I}_{i+1}-(\ket{0}\bra{0})_i\otimes(\ket{0}\bra{0})_{i+1}$. It is easy
to see that each part of Eq.\eqref{H3local} individually commutes with $O_i(=\ket{+}\bra{-}+\ket{0}\bra{0}+\ket{-}\bra{+})$.
Hence, $[H^{II},O_i]=0,\forall i$.

\section{Forward Scattering Approximation}
\label{appD}

\label{FSA}
Forward scattering approximation (FSA) has been an important tool to analyze the scar induced oscillation since
its first usage in spin-1/2 PXP model\cite{abanin1}. Here we will apply FSA in detail in our spin-1 models \cite{Bull}. 
The essential idea is to first break the model Hamiltonians in $H^+$ and $H^-$ (one is conjugate 
transpose to other) such that one (lets say $H^-$) annihilate the state $\ket{\mathbb{Z}_2}$ and the other ($H^+$) annihilate 
$\ket{\bar{\mathbb{Z}}_2}$. One state ($\ket{\bar{\mathbb{Z}}_2}$ / $\ket{\mathbb{Z}_2}$ ) can be obtained from the other
($\ket{\mathbb{Z}_2}$ / $\ket{\bar{\mathbb{Z}}_2}$) by repeated action ($2L$ times) of ($H^+$ / $H^-$). The oscillatory dynamics
then can be visualized as a coherent forward and backward scattering in between these two states. 

We define the n'th FSA vector ($\ket{v_n})$) by $\ket{v_n}=\frac{1}{\beta_n}H^+\ket{v_{n-1}}$ where 
$\ket{v_0}=\ket{\mathbb{Z}_2}$ and $\beta_n$ is the 
normalization constant. Due to the choice of the initial state and structure of the Hamiltonian, FSA vectors form a 
closed orthonormal subspace of dimension $2L+1$. Representation of the Hamiltonian in this subspace forms a tridiagonal 
matrix : $H_{FSA}=\beta_n\sum_{n=1}^{2L}\ket{v_n}\bra{v_{n+1}}+h.c$. Thus, if one is interested only in the scar subspace,
an enormous simplification can be achieved, namely, one need to diagonalize a matrix whose dimension scales only linearly with 
system size. This enables to deal with larger system size. In Fig. \ref{FSA1} we have compared the $\mathbb{Z}_2$-overlap of the 
eigenstates of $H_{FSA}$ and fidelity dynamics of the $\mathbb{Z}_2$ state generated by $H_{FSA}$ with the same 
quantities obtained by the full Hamiltonian ($H^I$) for Model-I. 
\begin{figure}[!]
\vskip 0.5cm
 \includegraphics[width=0.5\linewidth]{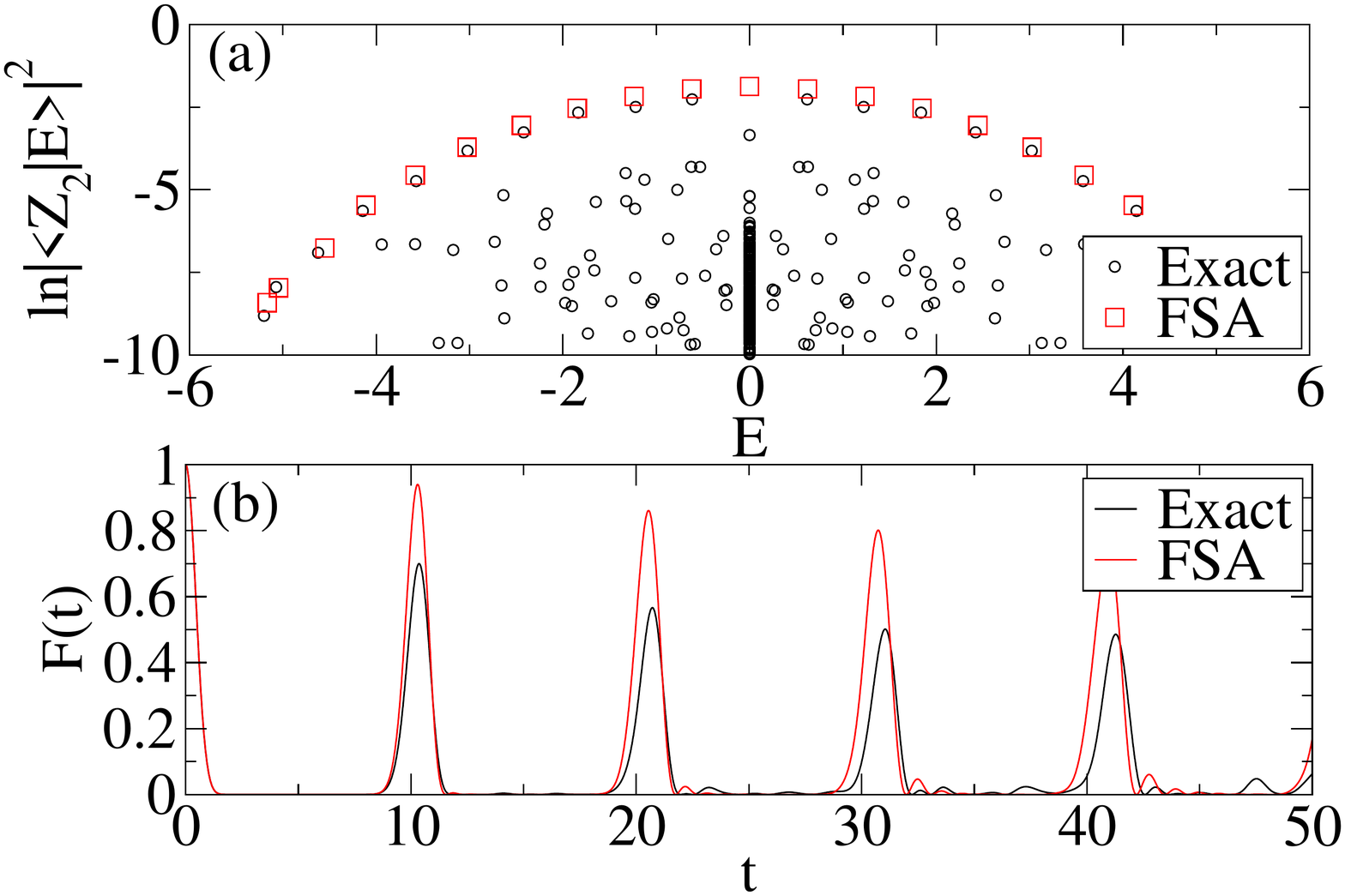}
 \caption{Comparison of FSA with exact numerics for Model-I (a) $\mathbb{Z}_2$-overlap (b) Fidelity dynamics of the $\mathbb{Z}_2$
 state. L=10.}
 \label{FSA1}
\end{figure}
To calculate other quantities (e.g. some observables, 
entanglement entropy etc) one needs to store the FSA vectors which again consumes exponential memory (though this scales as
$\sim Le^L$ but not as $\sim e^{2L}$.). The mismatch of the FSA results with exact numerics in Fig. \ref{FSA1} is due to the 
fact that the FSA vectors does not form a complete set and hence can't span the Hilbert space, that's why the leakage of the 
$\mathbb{Z}_2$ dynamics out of the scar (/FSA) manifold is inevitable. This causes the revival amplitude to decrease and 
dephase the dynamics. This can be quantified using the FSA errors : $\delta_n=||H^-\ket{v_n}-\beta_n\ket{v_{n-1}}||$. If the 
action of $H^-$ on a FSA vector completely undo the action of $H^+$ on the same vector, $\delta$ is zero and the corresponding
FSA step is exact. In an ideal paramagnet (described by $H=\sum_iS^x_i$) all FSA steps are exact but due to the constraints 
induced by the projectors, FSA errors are nonzero in $PXP$ models. Interestingly suitable term can be added to the bare $PXP$
model which can reduce\cite{Floquet} the FSA errors even to zero\cite{SU2}. In this situation the $\mathbb{Z}_2$ dynamics 
remains totally confined in the scar manifold which results in an undamped oscillation up to very long time. This suggests
that the FSA errors produces some kind of frictional force in the system and hence the amplitude of oscillation (/degree of 
fidelity revival) should be inversely proportional to the total (summed over all the FSA steps) amount of FSA errors present 
in the system. Motivated by these arguments, we plan to perform a detail analysis of the FSA errors for Model-I\textendash III.

\subsection{FSA for Model-I}
In Model-I $\ket{00}$,$\ket{+0}$,$\ket{0+}$ type of configurations are not allowed. We start with
\begin{equation}
 H^+_{I}\ket{v_0}=\frac{1}{\sqrt{2}}\sum_{i \ even}\ket{\cdots\mathop{0}_i\cdots}
\end{equation}
where $\cdots$ represents sea of repeated $+-/-+$ configurations (with a suitably added $+$ or $-$ at the end). 
So, $\beta_1=\sqrt{||H^+_{I}\ket{v_0}||}=\sqrt{L}/2$.
\begin{equation}
 \ket{v_1}=\frac{1}{\beta_1}H^+_{I}\ket{v_0}=\sqrt{\frac{2}{L}}\sum_{i \ even}\ket{\cdots\mathop{0}_i\cdots}
\end{equation}
One can easily check : $H^-_{I}\ket{v_1}=\beta_1\ket{v_0}$. Therefore, $\delta^{I}_1=0$.\\
Next,
\begin{eqnarray}
 H^+_{I}\ket{v_1}=\frac{1}{\sqrt{L}}(\sum_{i \ even}\ket{\cdots-\mathop{-}_i-\cdots}+
 2\sum_{\substack{i, j\ even\\ i\neq j}}\ket{\cdots\mathop{0}_i\cdots\mathop{0}_j\cdots})
\end{eqnarray}
There are $\frac{L}{4}(\frac{L}{2}-1)$ distinct 2nd type of states. 
The 2nd summation should be understood as a sum over only this many 
number of states (that's why we bring a factor 2 before it). All such summation notations in this paper avoid
double counting. We get,
$\beta_2=\sqrt{||H^+_{I}\ket{v_1}||}=\sqrt{\frac{1}{L}.\frac{L}{2}+\frac{4}{L}.\frac{L}{4}.(\frac{L}{2}-1)}=
\frac{\sqrt{L-1}}{\sqrt{2}}$.
and $\ket{v_2}=\frac{1}{\beta_2}H^+_{I}\ket{v_1}$. It is easy to see that $H^-_{I}\ket{v_2}=\beta_2\ket{v_1}$. 
$\therefore$ $\delta^{I}_2=0$.

Next we find
\begin{eqnarray}
 H^+_{I}\ket{v_2}=\sqrt{\frac{1}{L(L-1)}}(3\sum_{\substack{i,j \ even \\ i \neq j}}\ket{\cdots \mathop{0}_i
 \cdots -\mathop{-}_j- \cdots}+6\sum_{\substack{i,j,k \ even \\ i\neq j \neq k}}
 \ket{\cdots\mathop{0}_i\cdots\mathop{0}_j\cdots\mathop{0}_k\cdots})
\end{eqnarray}
the number of distinct 2nd type of states is 
$\frac{1}{3!}\frac{L}{2}(\frac{L}{2}-1)(\frac{L}{2}-2)=\frac{L(L-2)(L-4)}{48}$. Therefore, 
$\beta_3^2=\frac{1}{L(L-1)}(9.\frac{L}{2}.(\frac{L}{2}-1)+36.\frac{L(L-2)(L-4)}{48})=\frac{3(L-2)}{4}$.
$\ket{v_3}=\frac{1}{\beta_3}H^+_{I}\ket{v_2}$ and we again find $H^-_{I}\ket{v_3}=\beta_3\ket{v_2}$ which means
$\delta^3_{I}=0$.

Next we find
 \begin{eqnarray}
  H^+_{I}\ket{v_3}=\frac{\sqrt{2}}{\sqrt{3L(L-1)(L-2)}}(3\sum_{\substack{i,j \ even\\i\neq j}}
  \ket{\cdots-\mathop{-}_i-\cdots}+12\sum_{\substack{i,j,k \ even\\i\neq j \neq k}}\ket{\cdots\mathop{0}_i\cdots
  \mathop{0}_j\cdots-\mathop{-}_k-\cdots}+\nonumber\\24\sum_{\substack{i,j,k,l \ even\\i\neq j\neq k\neq l}}
  \ket{\cdots\mathop{0}_i\cdots\mathop{0}_j\cdots\mathop{0}_k\cdots\mathop{0}_l\cdots})
  \label{v4}
 \end{eqnarray}
Therefore, $\beta_4^2=||H^+_{I}\ket{v_3}||=\frac{2}{3L(L-1)(L-2)}(36.\frac{L}{4}(\frac{L}{2}-1)+144.\frac{L(L-2)(L-4)}{16}+
576.\frac{L(L-2)(L-4)(L-6)}{384})=L-3$
where we have used the fact that the number of distinct 2nd and 3rd type of states in Eq.\eqref{v4} are 
$\frac{L(L-2)(L-4)}{16}$ and $\frac{L(L-2)(L-4)(L-6)}{384}$ respectively. $\ket{v_4}=\frac{1}{\beta_4}H^+_{I}\ket{v_3}$
and one can again check $\delta^{I}_4=0$.Note that the first part of $H^+_{I}$ (i.e $\sum_{i \ odd}\sigma^+_i$) has 
null effect till now. 

Finally we arrive at the 5th step where non-zero FSA error arises for the first time. Calculations of FSA vectors
become cumbersome from this step onward as both part of $H^+_{I}$ will now have non-zero actions. After regrouping all
similar type of states we write a consolidated expression of the action of $H^+_I$ on $\ket{v_4}$
 \begin{eqnarray}
  H^+_{I}\ket{v_4}=\frac{1}{\sqrt{3L(L-1)(L-2)(L-3)}}(30\sum_{\substack{i,j,k \ even \\ i\neq j \neq k}}
  \ket{\cdots\mathop{0}_i\cdots-\mathop{-}_j-\cdots-\mathop{-}_k-\cdots}+
  6\sum_{i \ odd}\ket{\cdots--\mathop{0}_i--\cdots}+\nonumber\\
  60\sum_{\substack{i,j,k,l \ even\\i\neq j\neq k\neq l}}
  \ket{\cdots\mathop{0}_i\cdots\mathop{0}_j\cdots\mathop{0}_k\cdots-\mathop{-}_l-\cdots}+
  120\sum_{\substack{i,j,k,l,m \ even\\i\neq j\neq k\neq l\neq m}}\ket{\cdots\mathop{0}_i\cdots\mathop{0}_j\cdots 
  \mathop{0}_k\cdots\mathop{0}_l\cdots\mathop{0}_m\cdots})
   \label{v5}
 \end{eqnarray}
Counting the respective number of different states in Eq.\eqref{v5} we get
 \begin{eqnarray}
\beta_5^2&=&||H^+_{I}\ket{v_4}||\nonumber\\
&=&\frac{1}{3L(L-1)(L-2)(L-3)}[30^2.\frac{L(L-2)(L-4)}{16}+36.\frac{L}{2}+60^2.\frac{L(L-2)(L-4)(L-6)}{96}\nonumber\\
&&+120^2.\frac{L(L-2)(L-4)(L-6)(L-8)}{3840}]\nonumber\\
&=&\frac{5L^4-50L^3+175L^2-250L+144}{4(L-1)(L-2)(L-3)}
 \end{eqnarray}
The action of $H^-_{I}$ on $\ket{v_5}(=\frac{1}{\beta_5}H^+_{I}\ket{v_4})$ gives after grouping same type of states together
 \begin{eqnarray}
H^{-}_{I}\ket{v_5}&=&\frac{1}{\beta_5\sqrt{6L(L-1)(L-2)(L-3)}}[15(L-4)\sum_{\substack{i,j \ even\\i\neq j}}
\ket{\cdots-\mathop{-}_i-+-\cdots-\mathop{-}_j-\cdots}+\nonumber\\
&&(15L-54)\sum_{i \ odd}\ket{\cdots--\mathop{-}_i--\cdots}+
30(L-4)\sum_{\substack{i,j,k \ even\\i\neq j\neq k}}\ket{\cdots\mathop{0}_i\cdots\mathop{0}_j\cdots-\mathop{-}_k-\cdots}+\nonumber\\
&&60(L-4)\sum_{\substack{i,j,k,l \ even\\i\neq j\neq k\neq l}}\ket{\cdots\mathop{0}_i\cdots\mathop{0}_j\cdots\mathop{0}_k\cdots
\mathop{0}_l\cdots}]
\label{Hminusv5}
 \end{eqnarray}
note that the 1st and 2nd type of states in Eq.\eqref{Hminusv5} had same strength in $\ket{v_4}$ which is sufficient 
to see that $H^-_{I}\ket{v_5}$ is not proportional to $\ket{v_4}$. Thus finally error arises in 5th FSA step which can be easily 
calculated by evaluating the norm : 
$\delta^I_5=||H^-_{I}\ket{v_5}-\beta_5\ket{v_4}||=\frac{12(L^3-6L^2+11L-18)}{(L-1)(L-2)(L-3)(5L^4-50L^3+175L^2-250L+144)}$.

\subsection{FSA for Model-II}
Only $\ket{00}$ type of configurations are not allowed in Model-II. We start by
\begin{eqnarray}
 H^+_{II}\ket{\mathbb{Z}_2}&=&\frac{1}{\sqrt{2}}(\sum_{i \ even}\ket{\cdots-\mathop{0}_i-\cdots}+
 \sum_{i\ odd}\ket{\cdots-+\mathop{0}_i+-\cdots})
\end{eqnarray}
So $\beta_1=\sqrt{||H^+_{II}\ket{\mathbb{Z}_2}||}=\sqrt{L/2}$ and $\ket{v_1}=\frac{1}{\beta_1}H^+_{II}\ket{\mathbb{Z}_2}$. 
It is easy to see that $H^-_{II}\ket{v_1}=\beta_1\ket{\mathbb{Z}_2}$ and hence $\delta^{II}_1=0$. 
Next, 
\begin{eqnarray}
 H^+_{II}\ket{v_1}&=&\frac{1}{\sqrt{2L}}[\sum_{i \ even}\ket{\cdots-\mathop{-}_i-\cdots}+2\sum_{\substack{i,j \ even \\i\neq j}}
 \ket{\cdots-\mathop{0}_i-\cdots-\mathop{0}_j-\cdots}+2\sum_{\substack{i \ even \\ j \ odd}}\ket{\cdots-\mathop{0}_i-\cdots
 +\mathop{0}_j+\cdots}+\nonumber\\
 &&2\sum_{\substack{i,j \ odd \\ i\neq j}}\ket{\cdots+\mathop{0}_i+\cdots+\mathop{0}_j+\cdots}+
 \sum_{i \ odd}\ket{\cdots+\mathop{+}_i+\cdots}]
\end{eqnarray}
We get $\beta^2_2=||H^+_{II}\ket{v_1}||=\frac{2L-5}{2}$ and $\ket{v_2}=\frac{1}{\beta_2}H^+_{II}\ket{v_1}$. It is easy to check that $H^-_{II}\ket{v_2}=\beta_2\ket{v_1}$ which means $\delta^{II}_2=0$. Rarity of constraints produces 
a large number of states in the next step, we write the consolidated expression
\begin{eqnarray}
 H^+_{II}\ket{v_2}&=&\frac{1}{\sqrt{2L(2L-5)}}[3\sum_{\substack{i \ even \\ j \ odd}}\ket{\cdots-\mathop{-}_i-
 \cdots+\mathop{0}_j+\cdots}+\sum_{i \ even}(\ket{\cdots-\mathop{-}_i0\cdots}+\ket{\cdots0\mathop{-}_i-\cdots})+\nonumber\\
&& 3\sum_{i,j \ even}\ket{\cdots-\mathop{-}_i-\cdots-\mathop{0}_j-\cdots}+3\sum_{i,j \ odd}\ket{\cdots+\mathop{+}_i+\cdots
 +\mathop{0}_j+\cdots}+\sum_{i \ even}(\ket{\cdots+\mathop{+}_i0\cdots}+\ket{\cdots0\mathop{+}_i+\cdots})\nonumber\\
&& 3\sum_{\substack{i \ even \\ j \ odd}}\ket{\cdots-\mathop{0}_i-\cdots+\mathop{+}_j+\cdots}+
6\sum_{\substack{i,j,k \ even\\i\neq j\neq k}}\ket{\cdots-\mathop{0}_i-\cdots-\mathop{0}_j-\cdots-\mathop{0}_k-\cdots}+\nonumber\\
&&6\sum_{\substack{i,j,k \ odd\\i\neq j\neq k}}\ket{\cdots+\mathop{0}_i+\cdots+\mathop{0}_j+\cdots+\mathop{0}_k+\cdots}+
6\sum_{\substack{i,j \ even\\k \ odd}}\ket{\cdots-\mathop{0}_i-\cdots-\mathop{0}_j-\cdots+\mathop{0}_k+\cdots}\nonumber\\
&&6\sum_{\substack{i,j \ odd\\k \ even}}\ket{\cdots+\mathop{0}_i+\cdots+\mathop{0}_j+\cdots-\mathop{0}_k-\cdots}]
\label{model2FSA}
\end{eqnarray}
by carefully counting the number of each type of states in Eq.\eqref{model2FSA} we find 
\begin{equation}
 \beta_3^2=\frac{1}{2L(2L-5)}[2.(9.\frac{L}{2}.(\frac{L}{2}-2)+L+9.\frac{L}{2}.(\frac{L}{2}-2))+2.(36.\frac{L(L-2)(L-4)}{48}+
 36.\frac{L(L-4)(L-6)}{16})]=\frac{6L^2-45L+95}{4L-10}
\end{equation}
and $\ket{v_3}=\frac{1}{\beta_3}H^+_{II}\ket{v_2}$. We now calculate $H^-_{II}\ket{v_3}$
\begin{eqnarray}
 H^-_{II}\ket{v_3}&=&\frac{1}{\sqrt{2}}.\frac{1}{\sqrt{2L(2L-5)}}.\frac{1}{\beta_3}[(3.(\frac{L}{2}-2)+2+3.(\frac{L}{2}-1))
 (\sum_{i \ even}\ket{\cdots-\mathop{-}_i-\cdots}+\sum_{i \ odd}\ket{\cdots+\mathop{+}_i+\cdots})+\nonumber\\
 &&(6+6.\frac{L-6}{2}+6.\frac{L-6}{2})\sum_{\substack{i \ even \\j \ odd}}\ket{\cdots-\mathop{0}_i-\cdots+\mathop{0}_j+\cdots}
 +\nonumber\\
 &&(6+6.\frac{L-4}{2}+6.\frac{L-8}{2})(\sum_{i,j \ even}\ket{\cdots-\mathop{0}_i-\cdots+-+\cdots-\mathop{0}_j-\cdots}+
 \sum_{i,j \ odd}\ket{\cdots+\mathop{0}_i+\cdots+-+\cdots+\mathop{0}_j+\cdots})\nonumber\\
 &&(6+6.\frac{L-4}{2}+6.\frac{L-6}{2})\sum_{i \ even}(\ket{\cdots-\mathop{0}_i-0-\cdots}+\ket{\cdots+\mathop{0}_i+0+\cdots})]
\end{eqnarray}
we then find that $\delta^{II}_3=||H^-_{II}\ket{v_3}-\beta_3\ket{v_2}||=\frac{50(2L-9)}{(2L-5)(6L^2-45L+95)}$.

\subsection{FSA for Model-III}
In Model-III the $\ket{00}$ and $\ket{++}$ type of configurations are forbidden but $\ket{+}$ and $\ket{0}$ can sit next to 
each other. We start by
\begin{eqnarray}
 H^+_{III}\ket{\mathbb{Z}_2}&=&\frac{1}{\sqrt{2}}(\sum_{i \ even}\ket{\cdots-\mathop{0}_i-\cdots}+
 \sum_{i\ odd}\ket{\cdots-+\mathop{0}_i+-\cdots})
\end{eqnarray}
So $\beta_1=\sqrt{||H^+_{III}\ket{\mathbb{Z}_2}||}=\sqrt{L/2}$ and $\ket{v_1}=\frac{1}{\beta_1}H^+_{III}\ket{\mathbb{Z}_2}$. 
We checked that $H^-_{III}\ket{v_1}=\beta_1
\ket{\mathbb{Z}_2}$ and hence $\delta^{III}_1=0$.

Next 
\begin{eqnarray}
 H^+_{III}\ket{v_1}&=&\frac{1}{\sqrt{2L}}(\sum_{i \ even}\ket{\cdots-\mathop{-}_i-\cdots}+2\sum_{\substack{i,j \ even\\i\neq j}}
 \ket{\cdots-\mathop{0}_i-\cdots-\mathop{0}_j-\cdots}+2\sum_{\substack{i\ even\\j \ odd}}\ket{\cdots-\mathop{0}_i-\cdots
 +\mathop{0}_j+\cdots}+\nonumber\\
 &&2\sum_{\substack{i,j \ odd\\i\neq j}}\ket{\cdots+\mathop{0}_i+\cdots+\mathop{0}_j+\cdots})
 \label{model1FSA1}
\end{eqnarray}
counting distinct number of 4 different type of states in Eq.\eqref{model1FSA1}, we get 
$\beta_2^2=\frac{1}{2L}[\frac{L}{2}+4.\frac{L}{4}(\frac{L}{2}-1)+4.\frac{L}{2}(\frac{L}{2}-2)+4.\frac{L}{4}(\frac{L}{2}-1)]
=L-\frac{11}{4}$. $\ket{v_2}=\frac{1}{\beta_2}H^+_{III}\ket{v_1}$.

Next we find
\begin{eqnarray}
 H^-_{III}\ket{v_2}&=&\frac{1}{\sqrt{L(4L-11)}}[(2L-5)\sum_{i \ even}\ket{\cdots-\mathop{0}_i-\cdots}+
 (2L-6)\sum_{i \ odd}\ket{\cdots +\mathop{0}_i+\cdots}]
\end{eqnarray}
clearly $H^-_{III}\ket{v_2}$ is not equal to $\ket{v_1}$ multiplied by $\beta_2$ and this produces error in the 2nd FSA step. The 
norm of the difference is given by
\begin{eqnarray}
\delta^{III}_2=||H^-_{III}\ket{v_2}-\beta_2\ket{v_1}||=\frac{1}{4(4L-11)}
\end{eqnarray}

\end{widetext}
\end{document}